\documentclass[3p]{elsarticle}
\usepackage{hhline}
\usepackage{amsmath}
\usepackage{floatflt}

\newcommand{\be}{\begin{equation}}
\newcommand{\ee}{\end{equation}}
\newcommand{\bea}{\begin{eqnarray}}
\newcommand{\eea}{\end{eqnarray}}
\newcommand{\p}{\partial}

\newcommand{\rT}{\tilde r}
\newcommand{\hT}{\tilde h}
\newcommand{\zT}{\tilde z}
\newcommand{\TT}{\tilde T}
\newcommand{\tT}{\tilde t}
\newcommand{\urT}{\tilde u_r}
\newcommand{\uzT}{\tilde u_z}
\newcommand{\JT}{\tilde J}
\newcommand*{\Scale}[2][4]{\scalebox{#1}{\ensuremath{#2}}}%

\begin{document}
\begin{frontmatter}
\title{Marangoni convection in an evaporating droplet: Analytical and numerical descriptions}
\author[ITP,MIPT,SCC]{L.Yu. Barash}
\address[ITP]{Landau Institute for Theoretical Physics, 142432 Chernogolovka, Russia}
\address[MIPT]{Moscow Institute of Physics and Technology, 141700 Dolgoprudny, Russia}
\address[SCC]{Science Center in Chernogolovka, 142432 Chernogolovka, Russia }
\ead{barash@itp.ac.ru}
\begin{abstract}
The stationary single vortex Marangoni convection in an axially symmetrical sessile drop 
of capillary size is considered.
The detailed description of the fluid flows is presented for a wide range of contact angles, 
which takes into account the boundary conditions and the mass balance equation, 
without explicitly solving the Navier--Stokes equations.
The analytical approach developed is compared with the results of numerical simulations 
and demonstrated to describe reasonably well the single-vortex Marangoni flows.
This indicates the substantial role of the boundary conditions in the problem.
\end{abstract}
\end{frontmatter}

\section{Introduction}

The evaporation of a liquid droplet was studied since Maxwell time~\cite{Maxwell1877,Langmuir1918,Fuchs},
and has attracted much attention over the last decade
and a half in view of its role in various engineering applications,
the advent of nanotechnology and progress in understanding of
the evaporation process. In particular, the structure of the 
fluid flow produced by surface-tension-driven (Marangoni) instability inside
an evaporating droplet has been intensively 
studied~(see, for example,~\cite{Erbil,Larson} and references therein).

While numerical calculations of Marangoni convection
agree well with corresponding experimental data~\cite{HuLarsonReverse,SavinoFico,Thokchom},
existing analytical studies of capillary flows in droplets 
are either limited to the case of small contact angles (see Sec.~\ref{LubriSec}) 
or disregard Marangoni stresses at the droplet free surface~\cite{Masoud,Tarasevich,Gelderblom}.
Under the former conditions the problem is known to simplify and to be treated usually within 
the lubrication approximation. However, even in this case the analytical analysis
of the Marangoni convection has been restricted up to now by its combination with a 
numerical fitting of the temperature distribution over a free droplet surface, 
which plays a key role as a source of the Marangoni effect.
The latter approach can be valid when the Marangoni forces are suppressed
due to effects of surface surfactants or for other reasons.
However, generally, both the evaporative capillary flows and buoyancy-driven convection
are much weaker than the Marangoni fluid flow for a droplet 
of capillary size~\cite{HuLarsonMicrofluid,HuLarsonMarangoni,Barash2009},
and the consideration of surface tension gradients is necessary.
The primary purpose of this work is to study in detail the structure
of Marangoni convection in evaporating droplets with pinned contact line, 
with an emphasis on the effects of the boundary
conditions and the mass balance equations.  It is demonstrated that the fluid
flows can be obtained analytically for a wide range of contact angles, 
when the existing model of the
droplet evaporation, known mainly from numerical studies, is formulated
in a simplified manner. 

In order to calculate the fluid dynamics in an evaporating sessile droplet, one has
to solve numerically the coupled system of equations which contains 
the nonstationary vapor diffusion equation, the thermal conduction equation, 
the Navier--Stokes equations and to recalculate the
droplet shape at each step due to the evaporative mass loss for the respective time
interval~\cite{Barash2009}. 
The self-consistent solution should take into account 
an inhomogeneous evaporating flux density
in the boundary conditions for the thermal conduction equation,
since it is related to the heat transfer and 
hence to the temperature gradient at the droplet surface.
The variation of temperature over the droplet surface affects
the boundary conditions for the fluid dynamics,
since the surface tension depends on temperature.
In addition, the velocity field can influence the thermal conduction
as a result of the effects of heat convection.

\begin{figure}
\caption{The droplet and its element in the $rz$- and in the ${n\tau}$-coordinate system
correspondingly.}
\label{scheme}
\vspace{0.5cm}
\includegraphics[trim=0mm 3.1cm 0mm 0mm,clip,width=15cm]{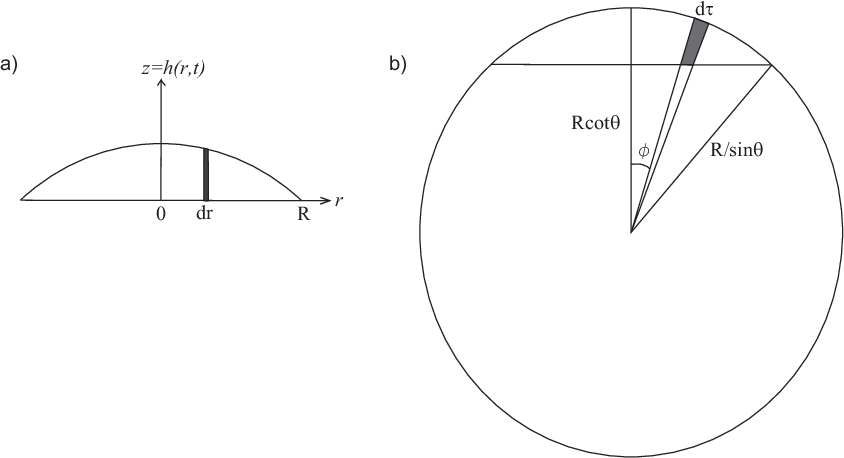}
\end{figure}

The calculations can be considerably simplified 
under the following conditions:
\begin{enumerate}
\item[a)] The capillary number ${\mathrm{Ca}}=\eta\overline{u}/\sigma$
and the Bond number ${\mathrm{Bo}}=\rho g h_0 R/(2\sigma\sin\theta)$
are much smaller than unity (see the notations in Table~\ref{ParamTable}). 
In this case, the sessile drop shape $h(r,t)$
can be approximated with high accuracy by the spherical 
cap approximation (see Fig.~\ref{scheme}b)
\begin{equation}
h(r,t)=\frac{R(\cos\phi(r,t)-\cos\theta(t))}{\sin\theta(t)};\qquad
\phi(r,t)=\arcsin\left(\frac{r\sin\theta(t)}{R}\right).
\label{hFromR}
\end{equation}
Here $\theta$ is the droplet contact angle,
$\overline{u}$ is the characteristic velocity and $h_0=h(0,t)$ is the
droplet height. The condition ${\mathrm{Ca}} \ll 1$ signifies that 
the viscous forces, which generally enter the boundary
condition for the pressure, are much smaller than capillary forces, 
and the hydrostatic Young--Laplace equation can be used in order to determine 
the droplet shape. Under the condition ${\mathrm{Bo}}\ll 1$,
influence of gravitational forces on the droplet shape is also small
(see, for example,~\cite{Barash2009}).

\item[b)] The inverse Stanton number ${\mathrm{St}}^{-1}=\overline{u}R/\kappa$
is much smaller than unity. In this case
the rate of the convective heat transfer is
much smaller than conductive heat transfer, and, hence, the velocity
field does not influence the thermal conduction~(see, for example,~\cite{HuLarsonMarangoni}).

\item[c)] The transient time for heat transfer $t_{heat}=Rh_0/\kappa$,
transient time for momentum transfer $t_{mom}=\rho Rh_0/\eta$ and
transient time for vapor phase mass transfer $t_{mass}=\rho_{vap}/\rho_f\cdot t_f$
should be much smaller than the total drying time $t_f\approx 0.2\rho R h_0/(D u_s)$.
This permits to describe the quasistationary stage of the evaporation process
disregarding the time derivatives in the heat conduction equation, 
Navier--Stokes equation and the diffusion equation.

\item[d)] The dimensionless number $\rho g h^2\beta/(7\sigma')$,
where $\beta$ is thermal expansion coefficient, is much smaller than unity. In this case
buoyancy-induced convection is much weaker than Marangoni flow~\cite{Pearson}.

\end{enumerate}

The laminar character of the flow should be also assumed.
A turbulent regime arises in the droplet
only for very large values of Marangoni and Reynolds
numbers~\cite{DuanBadam}. Therefore, the laminar character of the flow
does not imply that the Reynolds number ${\mathrm{Re}}=\rho \overline{u}R/\eta$
is much smaller than unity. The condition b) specified above
assumes that ${\mathrm{Re}}\cdot \nu/\kappa={\mathrm{St}}^{-1}$ is small.
Since Prandtl number is greater than one for the majority of liquids,
it follows from the condition that the Reynolds number is also assumed to be small.
Nevertheless, most of the relations derived in the present paper are not directly
connected to the calculation of the temperature distribution, and, therefore,
are potentially applicable to the case of larger Reynolds numbers.

It is natural to start from the vapor diffusion equation, which
can be solved independently from the other equations.
We focus on so-called lens model of evaporation, which is consistent
with the liquid--vapor interface at equilibrium and the evaporation limited by the
diffusion of vapour into the surrounding gas.
While other evaporative regimes are known in which the
evaporation rate is controlled not by vapor diffusion
but by phase change/kinetics processes at the 
interface~(see, e.g.,~\cite{MurisicKondic}),
it is known that the diffusion-limited evaporation model
agrees well with experimental data for moderately volatile droplets 
of capillary size with pinned contact line 
at atmospheric pressure of air and ambient temperature 
(see detailed discussions in, e.g.,~\cite{Larson,Barash2009,HuLarsonEvap,MurisicKondic,Ristenpart,Semenov}).
In particular, an evaporation of a droplet deposited onto a heated substrate,
where the atmospheric convective transport should be taken into account
together with the diffusive model~\cite{Sobac,Carle,Girard2011}, 
is beyond the scope of the present paper.
The analytical solution to the stationary vapor diffusion equation
with appropriate boundary conditions gives
the inhomogeneous evaporation flux from the surface of the evaporating droplet 
of a spherical shape~\cite{Deegan2000} 
\begin{equation}
J_s(r)=\frac{Du_s}{R}\left(\frac{\sin\theta}2+
\sqrt{2}(x(r)+\cos\theta)^{3/2}
\int_0^\infty\frac{\cosh(\theta\tau)}{\cosh(\pi\tau)}\tau\tanh((\pi-\theta)\tau)
P_{-1/2+i\tau}(x(r)) d\tau\right),
\label{JDeegan}
\end{equation}
where $x(r)={\left(r^2\cos\theta/R^2+\sqrt{1-r^2\sin^2\theta/R^2}\right)}/{(1-r^2/R^2)}$
and $P_{-1/2+i\tau}(x)$ is the Legendre polynomial.
It was shown in~\cite{Deegan2000} that~(\ref{JDeegan}) can be approximated with high accuracy as
\begin{equation}
J_s(r)=J_0(\theta)(1-r^2/R^2)^{-\lambda(\theta)},
\label{JDeeganLambda}
\end{equation}
where $\lambda(\theta)=1/2-\theta/\pi$
and explicit expressions for $J_0(\theta)$ can be found in~\cite{HuLarsonEvap}.

Under the condition b) specified above,
the heat conduction equation inside the droplet can
be solved independently, provided that the evaporation flux, which is
connected to the temperature gradient at the droplet surface
through the boundary condition $\p T/\p n=-LJ_s(r)/k$, is determined
by~(\ref{JDeegan}). 
We obtain numerically the surface temperature distribution of the drop
under the time independent conditions, which
is justified by the quasistationarity of the evaporation process 
(see the condition c) above). 
We use a fitting procedure for a surface temperature
which will be explained in more detail in Sec.~\ref{AnalysisSec}.
The following relation fits the obtained
quasistationary temperature at the droplet surface
with high accuracy,
where the parameters $a$, $b$, $c$ and $\Delta T$ were placed 
in Table~\ref{TfitTable}:
\begin{equation}
\frac{T}{\Delta T}=a\left(\frac{r}{R}\right)^b+(1-a)\left(\frac{r}{R}\right)^2+c.
\label{Tfit}
\end{equation}
Here $\Delta T$ is the temperature difference between the apex of the droplet
and the substrate. The droplet and fluid properties in Table~\ref{ParamTable} were used
during the numerical simulation.

For small contact angles the surface temperature distribution
will be obtained analytically in Sec.\ref{LubriSec}.
This allows completely analytical description
of the Marangoni convection in the lubrication approximation
for an evaporating droplet, as distinct from the previous 
considerations in~\cite{HuLarsonMicrofluid,HuLarsonMarangoni}.

The next step is the calculation of the fluid velocities in the droplet,
where the obtained profile of the surface temperature 
enters the boundary condition at the droplet surface
through the corresponding surface tension.
It was found in~\cite{Deegan2000,HuLarsonMicrofluid,HuLarsonMarangoni}
that this problem allows an approximate analytical
description at least for the case of droplets with relatively 
small contact angles (see Sec.~\ref{LubriSec}).
The derivation of the analytical description of the fluid flow
employs only the boundary conditions and the mass balance equations,
without explicitly solving the Navier--Stokes equations. 

For the analysis and interpretation of simulation results
it is quite desirable to have a simplified description of the phenomenon,
which nevertheless would be in a good agreement with the numerical data.
In this paper we will generalize the existing
analytical descriptions and will compare in detail the analytical
and the numerical results for various droplets.
We will describe how to accurately deal with the boundary conditions
for this problem and how to obtain a number of additional conditions 
from the mass balance equations.
We will find that a reasonable description 
of the fluid flows in a wide range of contact angles can be
obtained taking into account the boundary conditions and 
the mass balance equations,
without explicitly solving the Navier--Stokes equations.
Instead of solving the Navier--Stokes equation, which is not known to have
an analytical solution for such problem, we will use an appropriate approximation.
The conditions a) -- d) above are generally valid for a capillary size
liquid drop, except that strong Marangoni flows in relatively large volatile droplets
may transport heat rapidly enough that the convective heat transfer may not be negligible.
The results are applicable when convective heat transfer is negligible,
i.e. when ${\mathrm{St}}^{-1}=\overline{u}R/\kappa\ll 1$.
Also, we focus on pinned contact line configuration and we only consider $\theta<90^\circ$. 
A moving contact line or contact angles larger than $90^\circ$ could lead to additional issues 
not being considered in the present work.

Sec.~\ref{LubriSec} represents a brief review 
of the lubrication approximation derived in~\cite{HuLarsonMicrofluid,HuLarsonMarangoni},
completed with the analytical description of the corresponding surface temperature 
distibution and with further heuristic extension of the lubrication approximation.
The new analytical description resulting in better accuracy
is developed in Section~\ref{quadratic_ntau_Sec}.
Sec.~\ref{AnalysisSec} contains discussion of the obtained analytical and numerical results.

\section{The lubrication approximation}
\label{LubriSec}

The lubrication approximation for an evaporating droplet of capillary size was derived
in~\cite{HuLarsonMicrofluid,HuLarsonMarangoni}. The derivation includes
three basic assumptions which are justified for $\theta \ll 1$:
\begin{enumerate}
\item[a)] The radial velocity $u_r(z)$ at each value of $r$ has a quadratic dependence on $z$.
\item[b)] The droplet free surface is approximated as a parabola $h(\rT,t)=h(0,t)(1-\rT^2)$,
where $\rT=r/R$.
\item[c)] The total shear stress at the droplet free surface is approximated by the
$rz$-component of the stress tensor.
\end{enumerate}

Consider an axially symmetrical column in the droplet (see Fig.~\ref{scheme}a).
The mass balance equation states that the rate of mass change in the
given volume element is equal to the net flux of mass into the element:
\begin{equation}
\frac{d}{d t}(\delta m)=-\oint\rho\, {\mathbf u}\cdot d{\mathbf f},
\label{massbalance}
\end{equation}
where the vector $d\mathbf{f}$ is perpendicular to the surface of the element,
its absolute value is equal to the area of a small part of the boundary of the element,
and the amount of mass which evaporates each second from the surface of the element is assumed to be
also properly included in the surface integral in the right-hand side of~(\ref{massbalance}).
An easy consequence of the mass balance equation for the column was obtained in~\cite{Deegan2000}:
\begin{equation}
\overline{u}_r(r,t)=-\frac1{\rho r h}\int_0^r dr\, r\left(J_s(r,t)
\sqrt{1+\left(\frac{\p h}{\p r}\right)^2}+\rho\frac{\p h}{\p t}\right),
\label{ur_average}
\end{equation}
where $\overline{u}_r(r,t)$ is a height-averaged velocity in the column.
Let $\rT=r/R$, $\zT=z/h_0$, $\urT=u_r t_f/R$, $\uzT=u_z t_f/h_0$, $\tT=t/t_f$,
$\hT=h(r)/h_0$, $M_a=-\sigma'_T\Delta T t_f/(\eta R)$, $\JT=-J_0/\rho{\dot h_0}$.
Substituting in~(\ref{ur_average}) the free surface of the droplet as a parabola
$h(r,t)=h_0(t)(1-\rT^2)$ and the corresponding relations
$dh_0/d t=2(dm/dt)/(\rho\pi R^2)\approx h_0/(t-t_f)$, $dm/dt=\pi J_0 R^2/(1-\lambda)$,
Hu and Larson obtained in~\cite{HuLarsonMicrofluid} the relation
\begin{equation}
\frac{\overline{u}_r t_f}{R}=\frac{1}{4\rT}\frac{1}{1-\tilde t}
\left((1-\rT^2)^{-\lambda}-(1-\rT^2)\right).
\label{HuL_ur_average}
\end{equation}
This relation for the height-averaged velocity in evaporating droplets with relatively
small contact angles is the basis for the lubrication approximation for the droplets.
Using the above assumptions a), b), c) and Eq.~(\ref{HuL_ur_average}),
Hu and Larson derived the main lubrication equation for $u_r$, which takes the form
\begin{equation}
\urT= \frac38 \frac1{1-\tT}\frac1\rT\left[(1-\rT^2)-(1-\rT^2)^{-\lambda}\right]
\left(\frac{\zT^2}{\hT^2}-2\frac{\zT}{\hT}\right)+
\left[\frac{\rT h_0^2\hT}{R^2}\left(\JT\lambda(1-\rT^2)^{-\lambda-1}+1\right)+
\frac{M_a h_0\hT}{2R}\frac{d\TT}{d\rT}\right]
\left(\frac{\zT}{\hT}-\frac32\frac{\zT^2}{\hT^2}\right)
\label{lubri_urT}
\end{equation}
and implies $\hT(\rT,\tT)=\hT_0(\tT)(1-\rT^2)$.
As seen, $\urT(\rT,\zT)$ is the quadratic function of $\zT$ that
can be considered as a result of expansion in powers of
small parameter $\zT$ for droplets with small contact angles.
Eq.~(\ref{lubri_urT}) is applicable for arbitrary function $T(r)$
and at the same time it coincides with Eq.~(28)
in~\cite{HuLarsonMarangoni} provided that the surface
temperature distribution is described by Eq.~(\ref{Tfit}).

The final step of the derivation in~\cite{HuLarsonMicrofluid,HuLarsonMarangoni} 
is to find $\uzT$ using the obtained $\urT$
and the continuity equation for the incompressible fluid.
Here we represent the result in the form
\begin{multline}
\uzT= \frac34 \frac1{1-\tT}\left(1+\lambda(1-\rT^2)^{-1-\lambda}\right)
\left(\frac{\zT^3}{3\hT^2}-\frac{\zT^2}{\hT}\right)-
\frac34 \frac1{1-\tT}\left[(1-\rT^2)-(1-\rT^2)^{-\lambda}\right]
\left(\frac{\zT^2}{2\hT^2}-\frac{\zT^3}{3\hT^3}\right)\frac1{\rT}\frac{\p\hT}{\p\rT}-\\
-\frac{h_0^2}{R^2}\left(\JT\lambda(1-\rT^2)^{-\lambda-1}+1\right)
\left(\zT^2-\frac{\zT^3}{\hT}\right)-
\frac{\rT^2h_0^2}{R^2}\JT\lambda(\lambda+1)(1-\rT^2)^{-2-\lambda}
\left(\zT^2-\frac{\zT^3}{\hT}\right)-\\
-\frac{\rT h_0^2}{R^2}\left(\JT\lambda(1-\rT^2)^{-\lambda-1}+1\right)\frac{\zT^3}{2\hT^2}\frac{\p\hT}{\p\rT}-
\frac{M_a h_0}{4R}\left(\frac{d^2\TT}{d\rT^2}+\frac{1}{\rT}\frac{d\TT}{d\rT}\right)
\left(\zT^2-\frac{\zT^3}{\hT}\right)-
\frac{M_a h_0}{4R}\frac{d\TT}{d\rT}\frac{\zT^3}{\hT^2}\frac{\p\hT}{\p\rT}.
\label{lubri_uzT}
\end{multline}
Substituting $\hT(\rT,\tT)=\hT_0(\tT)(1-\rT^2)$
and~(\ref{Tfit}) in Eq.~(\ref{lubri_uzT}),
one obtains Eq.~(29) in~\cite{HuLarsonMarangoni}.
The assumption $\hT(\rT,t)=\hT_0(t)(1-\rT^2)$
is inherent to the lubrication approximation
and, in particular, to Eq.~(\ref{HuL_ur_average})
above and to Eqs.~(28) and~(29) in~\cite{HuLarsonMarangoni}.

The temperature distribution can be obtained analytically
for $\theta\ll 1$. 
Indeed, the boundary conditions for the temperature distribution 
take the form $\p T/\p r=0$ for $r=0$;
$T=T_0$ for $z=0$; ${\p T}/{\p n}=-LJ_s(r)/k$
at the drop surface.
In particular, for $r=0$ we have
$\left(\p T/\p z\right)_{r=0}=-LJ_0/k=-\Delta T/h_0$,
therefore, $\Delta T=LJ_0h_0/k$.
For small contact angles one can approximate the
boundary condition at the surface as ${\p T}/{\p z}=-LJ_s(r)/k$.
Therefore, using (\ref{hFromR}) and (\ref{JDeeganLambda}), one obtains
the surface temperature distribution:
\be
\frac{T}{\Delta T}=\frac{T_0}{\Delta T}-\frac{LJ_s(r)h(r)}{k\Delta T}=\frac{T_0}{\Delta T}-\frac{J_s(r)h(r)}{J_0h_0}=
\frac{T_0}{\Delta T}-\left(1-\frac{r^2}{R^2}\right)^{1/2+\theta/\pi}.
\label{Tlubri}
\ee
For small contact angles the rate of mass loss is
$-dm/dt=-\rho\pi R^3\theta'(t)/4=2\pi J_0(\theta)R^2$,
hence one can estimate $J_0$ and $\Delta T$ as follows:
$J_0(\theta)=-\rho R \theta'(t)/8$ and $\Delta T=LJ_0h_0/k=-\rho R^2\theta\theta'(t)L/(16k)$.
Eq.~(\ref{Tlubri}) obtained here agrees well with the numerical simulations
for small contact angles.

The main purpose of this work is to develop analytical
description of fluid flows in an evaporating droplet 
in a wide range of contact angles.
As a first step, we will test the heuristic extension of the lubrication approximation, 
substituting in Eqs.~(\ref{lubri_urT}) and~(\ref{lubri_uzT})
the functions $\hT(\rT)$ and $\p\hT/\p\rT$, which are obtained from~(\ref{hFromR}),
i.e., which correspond to the spherical cap profile of the sessile drop.
The equations were written in the form of Eqs.~(\ref{lubri_urT}) and~(\ref{lubri_uzT}) 
in order to represent heuristic description for larger contact angles,
where $h(r)$ is substantially nonparabolic.
This improves the accuracy for larger contact angles,
without changing the solution for small contact angles.
We note that the continuity equation is precisely satisfied
for arbitrary $h(r)$ and $T(r)$, if $\urT$ and $\uzT$
are represented in the form~(\ref{lubri_urT}),~(\ref{lubri_uzT}),
as opposed to the original form of lubrication equations,
where the continuity equation is satisfied for small contact angles. 
We formulate and use the heuristic extension to larger contact angles
for comparison with more consistent results obtained in this work.

\section{Derivation of the description in the
$\Scale[1.5]{n\tau}$-coordinate system}
\label{quadratic_ntau_Sec}

In this section the analytical approach for calculating
the fluid velocities in an evaporating droplet will be 
consistently and explicitly developed 
without using the assumptions $a), b), c)$ of Section~\ref{LubriSec}. 
The boundary conditions at the droplet surface 
will be considered assuming its spherical profile.

Consider the spherical $n\tau$-coordinate system, 
where $n$ is the distance between a point inside the droplet
and the center of the sphere which contains the droplet surface,
$\phi$ is the azimuthal angle,
and $\tau=n\phi$ (see Fig.~\ref{scheme}b).
The notation was taken to emphasize that the coordinates correspond to
the normal and tangential directions to the droplet surface.
Therefore, at the substrate we have $n=n_1=R\cot\theta/\cos\phi$, and
at the droplet surface we have $n=n_2=R/\sin\theta$. The $n\tau$-coordinates
are connected with the cylindrical $rz$-coordinates via the following relations:
$r=n\sin\phi$; $z=n\cos\phi-R\cot\theta$;
$n=\sqrt{r^2+(z+R\cot\theta)^2}$; $\tau=n\phi$; $\phi=\arcsin(r/n)$.
Here $\theta$ is the contact angle, therefore $\phi\le\theta$
for arbitrary point inside the droplet.

The total mass of the shaded element in Fig.~\ref{scheme}b is
\begin{equation}
\delta m =\rho \int_{n_1}^{n_2} 2\pi r(n) d\tau(n) dn = 2\pi\rho \int_{n_1}^{n_2} (n\sin\phi)(nd\phi) dn
=\frac23\pi\rho\frac{R^3\sin\phi\cdot d\phi}{\sin^3\theta(t)} \left(1-\frac{\cos^3\theta(t)}{\cos^3\phi}\right).
\end{equation}
The volume element is characterized by a fixed value of $\phi$,
while the center of the sphere is moving in a downward direction 
during the evaporation process. Therefore, the element is also moving and
the velocity of the element is $-R\theta'(t)\sin\phi/\sin^2\theta$.
It follows from the mass balance equation~(\ref{massbalance}) for the element that
\begin{equation}
\frac{d}{dt}(\delta m) = -J_s(\tau,t)\cdot 2\pi (n_2\sin\phi)d\tau
-2\pi\rho\cdot d\phi\frac{d}{d\phi}\left(\sin\phi\int_{n_1}^{n_2}n
\left(u_\tau+\frac{R\theta'(t)\sin\phi}{\sin^2\theta}\right)dn\right),
\end{equation}
where $J_s$ can be obtained with~(\ref{JDeeganLambda}).
Hence one obtains the following relation for $I(\phi)=\int_{n_1}^{n_2}nu_\tau dn$:
\begin{equation}
\label{MassBalance}
I(\phi)=
\frac{-R^2}{\sin^2\theta\sin\phi} \int_0^\phi \left(
\frac{J_0}{\rho}\left(1-\frac{\sin^2\phi}{\sin^2\theta}\right)^{-\lambda}
+\frac{R\cos\theta\cdot\theta'(t)}{\sin^2\theta}\left(\frac{\cos\theta}{\cos^3\phi}-1\right)
\right)\sin\phi\cdot d\phi
-\frac{R^3\theta'(t)\sin\phi}{2\sin^4\theta}\left(1-\frac{\cos^2\theta}{\cos^2\phi}\right).
\end{equation}
Here one can use the approximation $\theta'(t)\approx \theta/t_f$,
where $t_f(t)$ is remaining time of evaporation. Indeed, the contact
angle diminishes almost linearly with time
during the evaporation process~\cite{Barash2009}.
Eq.~(\ref{MassBalance}) results in the singularity in $u_\tau$ at the contact
line, where $n_1(\phi)\to n_2$, which is a consequence of the singularity
in evaporation rate at the contact line. Similar singularity problems
are known for all known analytical models. 
In particular, the singularity enters Eqs.~(\ref{ur_average})--(\ref{lubri_uzT}).
The singularity can be regularized by introducing 
a disjoining pressure and/or precursor film~\cite{deGennes,Bonn},
by introducing a Navier slip~\cite{Huh,Bonn,PetsiBurganos2008}
or by taking into account the Kelvin effect~\cite{Rednikov,RednikovAPS}.
The singularity influences the velocity field only in a small vicinity
of the contact line and its detailed discussion is beyond the
scope of the present paper.

At the droplet surface we have the following boundary
condition (see Appendix B in~\cite{Barash2009}):
\begin{equation}
\label{MarangoniBoundary}
\frac{d\sigma}{\eta d\tau}=-\frac{M_a\cos\phi}{t_f}\frac{d\tilde T}{d\tilde r}
=\frac{\p u_\tau}{\p n}+\frac{\p u_n}{\p\tau} - u_\tau\frac{d\phi}{d\tau}.
\end{equation}
This very important boundary condition takes into account Marangoni
forces associated with the temperature dependence of the surface tension,
which generate fluid convection in the sessile drop.
Additional condition at the droplet surface can be derived from the continuity equation:
\begin{equation}
\frac{u_n}{\cos\phi}=u_r\tan\phi+u_z=-u_r h'(r)
-\frac{1}{r}\int_0^{h(r)}\frac{\p(ru_r)}{\p r}dz=
-\frac{1}{r}\frac{\p(rh\overline{u}_r)}{\p r},
\end{equation}
hence, using the relation~(\ref{ur_average}), one obtains at the droplet surface
\begin{equation}
\label{unEquation}
u_n=\frac{\p h(r,t)}{\p t}\cos\phi +
\frac{J_0(\theta)}{\rho}\left(1-\frac{\sin^2\phi}{\sin^2\theta}\right)^{-\lambda(\theta)}.
\end{equation}
Substituting~(\ref{hFromR}) in~(\ref{unEquation}), one gets
\begin{eqnarray}
\label{un}
u_n&=&\frac{R(\cos\phi-\cos\theta)\theta'(t)}{\sin^2\theta}
+\frac{J_0}{\rho}\left(1-\frac{\sin^2\phi}{\sin^2\theta}\right)^{-\lambda},\\
\frac{\p u_n}{\p\tau}&=& \frac{\sin\theta}{R}\frac{\p u_n}{\p\phi}
=\frac{\sin\phi}{R\sin\theta}\left(\frac{2J_0\lambda\cos\phi}{\rho}
\left(1-\frac{\sin^2\phi}{\sin^2\theta}\right)^{-\lambda-1}-R\theta'(t)\right).
\label{dundtau}
\end{eqnarray}
We will use the following approximation for $u_\tau$:
\begin{equation}
u_\tau(n,\phi)=(n-n_1)^{p(\phi)}A(\phi)+(n-n_1)B(\phi).
\label{quadratic}
\end{equation}
Here $p(\phi)$ is a trial function.
The approach reasonably works for various trial functions $p(\phi)$.
We will choose $p(\phi)$ in Section~\ref{AnalysisSec}.
The coefficient functions $A(\phi)$ and $B(\phi)$ will be specified
based on the boundary condition~(\ref{MarangoniBoundary})
and the mass balance relation~(\ref{MassBalance}).
Eq.~(\ref{quadratic}) automatically satisfies 
the no-slip boundary condition at the substrate
and, at the same time, allows the existence 
of a single vortex in a droplet.

Using~(\ref{quadratic}), Eq.~(\ref{MarangoniBoundary}) can be rewritten as
\begin{equation}
A(\phi)(n_2-n_1)^{p(\phi)-1}\left(p(\phi)-1+\frac{n_1}{n_2}\right)
+B(\phi)\frac{n_1}{n_2}+\frac{\p u_n}{\p\tau}+\frac{M_a\cos\phi}{t_f}\frac{d\tilde T}{d\tilde r}=0,
\label{FirstNTau}
\end{equation}
where the right side of (\ref{dundtau}) can be used instead of ${\p u_n}/{\p\tau}$.

Eq.~(\ref{MassBalance}) gives the second linear relationship between $A(\phi)$ and $B(\phi)$:
\begin{equation}
\frac{A(\phi)}{(p+1)(p+2)}(n_2-n_1)^{p+1}(n_2(p+1)+n_1)+\frac16 B(\phi)(n_2-n_1)^2(n_1+2n_2)=I(\phi).
\label{SecondNTau}
\end{equation}
We note that the integral in the right side of (\ref{MassBalance}) can be obtained exactly, because
\begin{equation}
\int_0^\phi \sin\phi\left(\frac{\cos\theta}{\cos^3\phi}-1\right)d\phi=\cos\phi-1+
\frac12\cos\theta\tan^2\phi,
\end{equation}
\begin{equation}
\int\sin\phi\left(1-\frac{\sin^2\phi}{\sin^2\theta}\right)^{-\lambda}d\phi=
(\cos\theta+\cos\phi)\left(1-\frac{\sin^2\phi}{\sin^2\theta}\right)^{-\lambda}
{}_2F_1\left(1,\lambda;2\lambda;\frac{2\cos\theta}{\cos\theta-\cos\phi}\right)
\frac{\Gamma(2\lambda-1)}{\Gamma(2\lambda)}+C,
\end{equation}
where ${}_2F_1(a,b;c;z)$ is the hypergeometric function and $\Gamma(z)$ is the gamma function.
Therefore,
\begin{multline}
I(\phi)=\frac{-R^2J_0}{\rho\sin^2\theta\sin\phi}
(\cos\theta+\cos\phi)\left(1-\frac{\sin^2\phi}{\sin^2\theta}\right)^{-\lambda}
{}_2F_1\left(1,\lambda;2\lambda;\frac{2\cos\theta}{\cos\theta-\cos\phi}\right)
\frac{\Gamma(2\lambda-1)}{\Gamma(2\lambda)}+\\+
\frac{R^2J_0(\cos\theta+1)}{\rho\sin^2\theta\sin\phi}
{}_2F_1\left(1,\lambda;2\lambda;\frac{2\cos\theta}{\cos\theta-1}\right)
\frac{\Gamma(2\lambda-1)}{\Gamma(2\lambda)}+
\frac{R^3\cos\theta\cdot\theta'(t)}{\sin^4\theta\sin\phi}
\left(1-\cos\phi-\frac{\sin^2\phi}{2\cos\theta}\right).
\label{integratedI}
\end{multline}

The quantity $I(\phi)$ is typically negligibly small. 
In particular, it is negligibly small for
all droplets considered in Table~\ref{ParamTable}
and $I(\phi)=0$ can be used in order to simplify the equations,
because the evaporative-driven flows and the flows 
due to the nonzero value of $\theta'(t)$ are tiny
compared to the flow due to the Marangoni forces.
The calculations confirm that for describing the Marangoni
convection, the value of $I(\phi)$ could be replaced by zero almost without
losing the accuracy. 
The quantity $I(\phi)$ and Eq.~(\ref{integratedI}) could play 
an important role for describing
the evaporative-driven flows in other situations, when the 
Marangoni forces are suppressed. For this reason, we will
retain the quantity $I$ in the formulae.
The calculations also show that the effect of nonzero $u_n$ and $\p u_n/\p\tau$
on the fluid flows (see Eqs.(\ref{un}) and (\ref{dundtau}))
is quite small compared to the effect of Marangoni forces.
Hence $\p u_n/\p\tau=0$ can be used in order to further simplify the equations.
Still, we will retain $\p u_n/\p\tau$ in the formulae, because
it could possibly be useful in situations involving
extremely volatile liquids and a very fast evaporation.

Eqs.~(\ref{FirstNTau}),(\ref{SecondNTau}) are the system of two linear relationships 
between $A(\phi)$ and $B(\phi)$. The solution is:
\begin{equation}
A= -\frac{(-n_1+n_2)^{-1-p} (1+p) (2+p) \left(\left(6 I n_1+\frac{\p u_n}{\p\tau} (n_1-n_2)^2 n_2 (n_1+2 n_2)\right) t_f+\frac{\p \tilde T}{\p\tilde r} M_a (n_1-n_2)^2 n_2 (n_1+2 n_2) \cos\phi\right)}{(p-1) \left(2 n_2^2 (p+1) (p+2)+n_1^2 (p+4)+n_1 n_2 (p+1) (p+4)\right) t_f},
\label{AfromPhi}
\end{equation}
\begin{equation}
B= \frac{6 \left(I (p+1) (p+2) (n_1+n_2(p-1)) t_f+(n_1-n_2)^2 n_2 (n_1+n_2 (p+1)) ((\p u_n/\p\tau) t_f+(\p\tilde T/\p\tilde r) M_a \cos\phi)\right)}{(n_1-n_2)^2 (p-1) \left(2 n_2^2 (p+1) (p+2)+n_1^2 (p+4)+n_1 n_2 (p+1) (p+4)\right) t_f}.
\label{BfromPhi}
\end{equation}
In order to complete the approximate analytical description of the velocity field, we
need to obtain $u_n(n,\phi)$ inside the droplet. The continuity equation for
the incompressible fluid $\mathrm{div}\,\mathbf{u}=0$
takes the following form in the $n\tau$-coordinate system:
\begin{equation}
\frac{\p u_n}{\p n}+\frac{2u_n}{n}+\frac{\p u_\tau}{\p\tau}+\frac{\cot\phi}{n}u_\tau=0.
\end{equation}
Therefore,
\begin{multline}
\frac{1}{n}\frac{\p(n^2 u_n(n,\phi))}{\p n}=-\frac{\p u_\tau}{\p\phi}-u_\tau\cot\phi =
-A'(\phi)(n-n_1)^{p(\phi)}-B'(\phi)(n-n_1)+\\+
B(\phi)(n_1'(\phi)-(n-n_1)\cot\phi)+
A(\phi)p(\phi)n_1'(\phi)(n-n_1)^{p(\phi)-1}-
A(\phi)(n-n_1)^{p(\phi)}(p'(\phi)\log(n-n_1)+\cot\phi),
\end{multline}
\begin{multline}
u_n(n,\phi)=-\frac{A'(\phi)(n-n_1)^{p+1}}{n^2}\frac{n_1+np+n}{(p+1)(p+2)}
-\frac{B'(\phi)}{6n^2}(n-n_1)^2(n_1+2n)+\frac{B(\phi)}{2n^2}(n^2-n_1^2)n_1'(\phi)-\\-
\frac{B(\phi)(2n+n_1)(n-n_1)^2\cot\phi}{6n^2}+
\frac{A(\phi)(pn+n_1)(n-n_1)^pn_1'(\phi)}{(p+1)n^2}-
\frac{A(\phi)\cot\phi}{(p+1)(p+2)}\frac{(n_1+np+n)(n-n_1)^{p+1}}{n^2}+\\+
\frac{A(\phi)p'(\phi)(n-n_1)^{p+1}}{n^2}\left(
\frac{n}{(p+2)^2}+\frac{n_1(2p+3)}{(p+1)^2(p+2)^2}-\frac{(n+n_1+np)\log(n-n_1)}{(p+1)(p+2)}
\right).
\label{u_n}
\end{multline}
Therefore, we have obtained the velocity field in the droplet:
$u_\tau$ and $u_n$ are defined by (\ref{quadratic}) and (\ref{u_n}), 
where $A(\phi)$ and $B(\phi)$
are determined by (\ref{AfromPhi}) and (\ref{BfromPhi}).
The velocities in the cylindrical $rz$-coordinate system can be obtained
from the known values of $u_n$ and $u_\tau$ with the following relations:
\begin{eqnarray}
u_r(r,z)&=&u_\tau(n,\phi)\cos\phi + u_n(n,\phi)\sin\phi,\\
u_z(r,z)&=&-u_\tau(n,\phi)\sin\phi + u_n(n,\phi)\cos\phi,
\end{eqnarray}
where $n=\sqrt{r^2+(z+R\cot\theta)^2}$ and $\phi=\arcsin(r/n)$.

There is also an analytical estimate for the surface temperature distribution.
Consider the circular arc intersecting orthogonally both the droplet surface
and the substrate. Its length is $d(r)=h(r)\phi/\sin\phi$.
Assuming a constant value of the temperature gradient along the arc, we obtain an
approximation for the surface temperature distribution:
\be
\frac{T}{\Delta T}=\frac{T_0}{\Delta T}-\frac{LJ_s(r)d(r)}{k\Delta T}=
\frac{T_0}{\Delta T}-\left(1-\frac{r^2}{R^2}\right)^{-1/2+\theta/\pi}\frac{\cos\phi-\cos\theta}{1-\cos\theta}\frac{\phi}{\sin\phi}.
\label{TAnalytical}
\ee
We find reasonably good agreement between the approximate relation (\ref{TAnalytical}) 
and our numerical results for the surface temperature distribution.
In the limiting case $\theta\ll 1$ Eq.(\ref{TAnalytical}) coincides with (\ref{Tlubri}).

\section{Numerical results and discussion}
\label{AnalysisSec}

\begin{figure}[tb]
\caption{ The function $p(\phi)$, which results in a best fit
between Eq.(\ref{quadratic}) and numerically obtained tangential
velocity. Here x-axis corresponds to $\phi/\theta$, where
$\theta=50^\circ$ is the droplet contact angle.
Red curve corresponds to the ethanol droplet,
blue curve corresponds to the hexanol droplet. 
Using the shown values of $p(\phi)$, the numerical plots of $u_\tau(n)$ for each $\phi$
are visually indistinguishable from their best fit with Eq.~(\ref{quadratic}).
Green curve shows the plot of Eq.~(\ref{TrialFunction}).
}
\label{TrialFunctionBestFit}
\includegraphics[width=8cm]{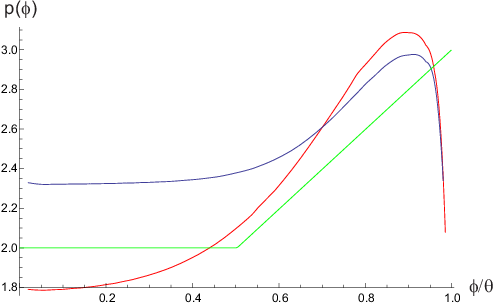}
\end{figure}

Table~\ref{ParamTable} shows the parameter values that were used in the calculations.
The fluid and vapor properties were taken from~\cite{Lide}.
Table~\ref{TfitTable} shows the contact angle, contact line radius and fitting parameters 
corresponding to Eq.~(\ref{Tfit}) for the 25 droplets.
For each droplet, we have $c=T_0/\Delta T-1$, and the value of $a$
is obtained with the least squares fit for a given values of $b$ and $c$.
The parameter $b$ is an integer which is chosen to give the minimal
value of the numerically obtained integral

\begin{equation}
\int_0^1 \left(\frac{T(\rT)}{\Delta T}-(a\rT^b+(1-a)\rT^2+c)\right)^2 d\rT,
\end{equation}
where the surface temperature distribution of the drop $T(\rT)$ is obtained numerically
under the time independent conditions.
The boundary conditions for the stationary equation $\Delta T=0$,
which is numerically solved inside the drop, take the form $\p T/\p r=0$ for $r=0$;
$T=T_0$ for $z=0$; ${\p T}/{\p n}=-LJ_s(r)/k$
at the drop surface. Here $\mathbf{n}$ is a normal vector to the drop surface,
$J_s$ is determined by~(\ref{JDeegan}), other notations are explained in Table~{\ref{ParamTable}}.
The results in Table~\ref{TfitTable} are obtained using the boundary condition $T=T_0$ for $z=0$,
which assumes high thermal conductivity of the substrate.
High thermal conductivity of the substrate, generally, results in a single-vortex fluid flow in the droplet,
while other convective regimes are also known, where the thermal conduction inside the substrate 
is important and should be taken into account~\cite{Ristenpart,BarashSubstrate1,BarashSubstrate2,Zhang}.

The numerical results show that the power exponent 
$p(\phi)$ in Eq.~(\ref{quadratic}) is, generally, between $1.5$ and $3.5$.
Some of such numerical results for $p(\phi)$ are shown 
in Fig.~\ref{TrialFunctionBestFit}.
The numerically obtained tangential velocities $u_\tau(n,\phi)$ 
are best fitted with Eq.~(\ref{quadratic}) with
the power exponents shown in Fig.~\ref{TrialFunctionBestFit},
where the red curve corresponds to the ethanol droplet and the blue curve
corresponds to the hexanol droplet.
Using these values of $p(\phi)$, the numerical plots of $u_\tau(n)$ for each $\phi$
are visually indistinguishable from their best fit with Eq.~(\ref{quadratic}).
Thus, using Eq.~(\ref{quadratic}) with $p(\phi)$ between $1.5$ and $3.5$
gives reasonably accurate description of $u_\tau(n)$ even 
for relatively complex single-vortex fluid flows.

Taking the trail function $p(\phi)=2$ already allows to achieve a reasonable accuracy 
in describing the velocity field in various droplets. In order to choose
the trial function closer to its actual behavior, such as shown in Fig.~\ref{TrialFunctionBestFit},
we have specified the trial function $p(\phi)$ to take the form
\begin{equation}
p(\phi)=
\left\{
\begin{aligned}
2 &,\mbox{\qquad for\quad} \phi/\theta \leq 1/2,  \\
2+{4\delta}/{\pi}\left(1-\cos\left({\pi}\left({\phi}/{\theta}-1/2\right)/{(2\delta)}\right)\right) &,  
\mbox{\qquad for\quad} 1/2 < \phi/\theta \leq 1/2+\delta,\\
1+2{\phi}/{\theta}-2(1-2/\pi)\delta &, \mbox{\qquad for\quad} \phi/\theta>1/2+\delta,
\end{aligned}
\right.
\label{TrialFunction}
\end{equation}
where $\delta=10^{-2}$.
Plot of Eq.~(\ref{TrialFunction}) is shown as a green curve in Fig.~\ref{TrialFunctionBestFit}.
This function is equal to $2$ for $\phi \leq \theta/2$,
it smoothly changes from $2$ to $3$ when $\phi$ changes from $\theta/2$ to $\theta$
and its derivative is continuous.
We note that the function $p(\phi)$ does not depend on the contact angle or the liquid properties.

\begin{figure}[t]
\caption{The surface velocity (cm/s) vs $r$ (cm) for the hexanol droplet with
a) $\theta=50^\circ$ and b) $\theta=20^\circ$.
Blue curve is the numerically obtained surface velocity. 
Green curve: the surface velocity in the heuristic extension of the lubrication approximation.
Purple curve: the surface velocity in the $n\tau$-description derived in Sec.~\ref{quadratic_ntau_Sec}.
Right panel: color scale for Figs.~\ref{h4t_vect} and~\ref{t4t_vect}, where
the values $u_{max}$ are given in Table~\ref{TabVeloc1}.
}
\label{hexanol_surface}
\includegraphics[width=0.39\textwidth]{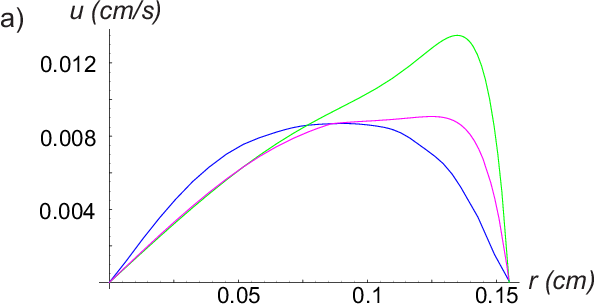}
\includegraphics[width=0.39\textwidth]{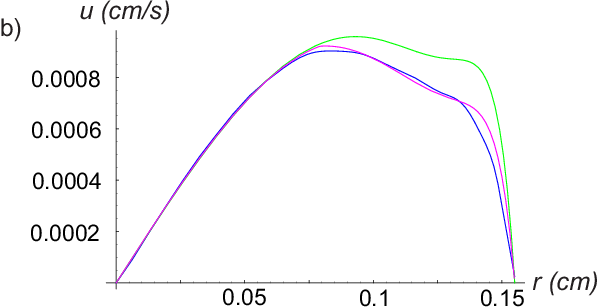}
\includegraphics[width=0.20\textwidth]{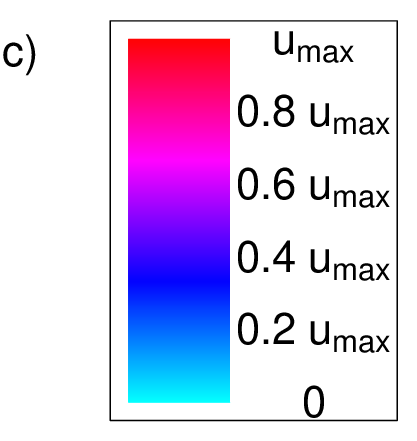}
\end{figure}
\begin{figure}
\caption{Vector field plots of the velocity field containing single vortex
for the hexanol droplet with $\theta=50^\circ$:
numerically obtained velocity field, the velocity field
in the heuristic extension of the lubrication approximation and in the $n\tau$-description.
}
\label{h4t_vect}
\vspace{0.2cm}
\includegraphics[width=5.3cm]{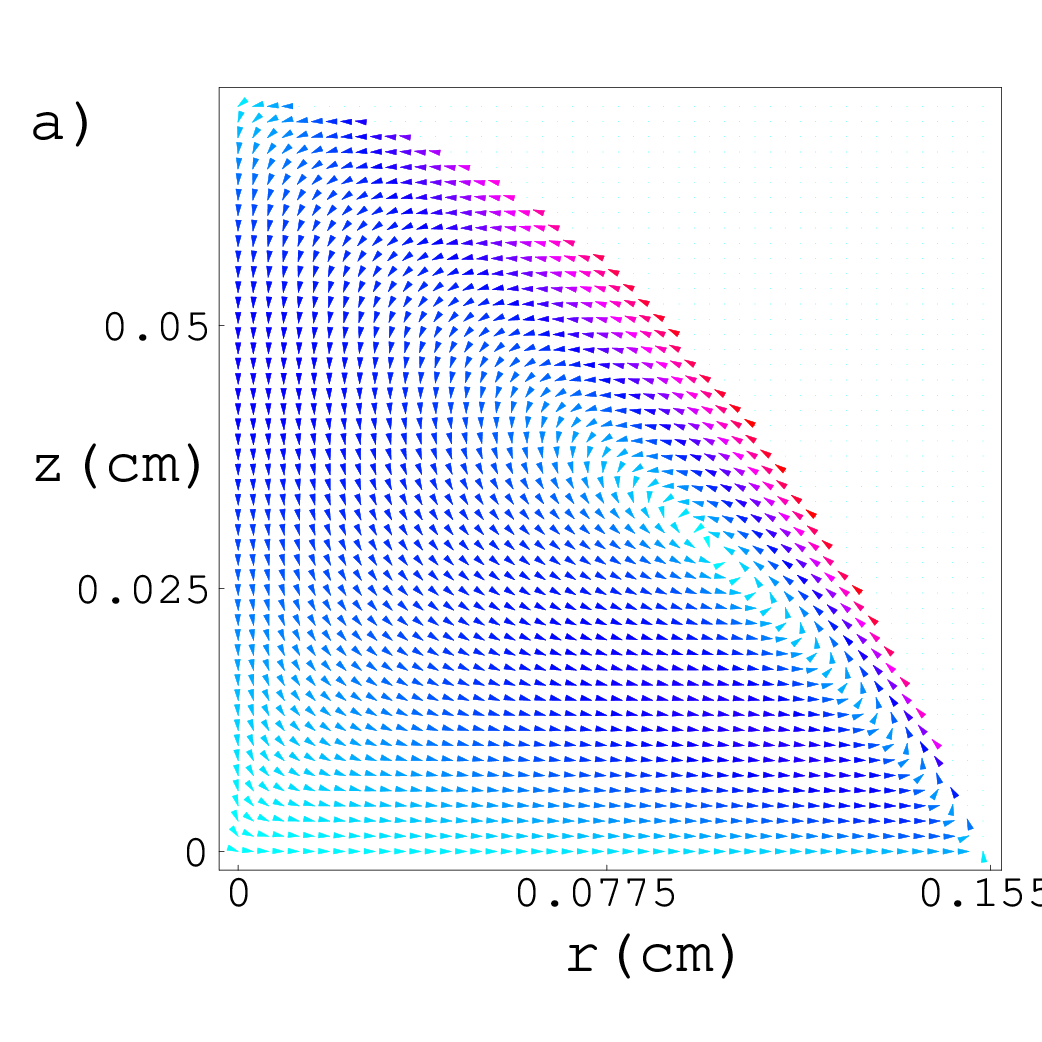}
\hspace{0.1cm}
\includegraphics[width=5.3cm]{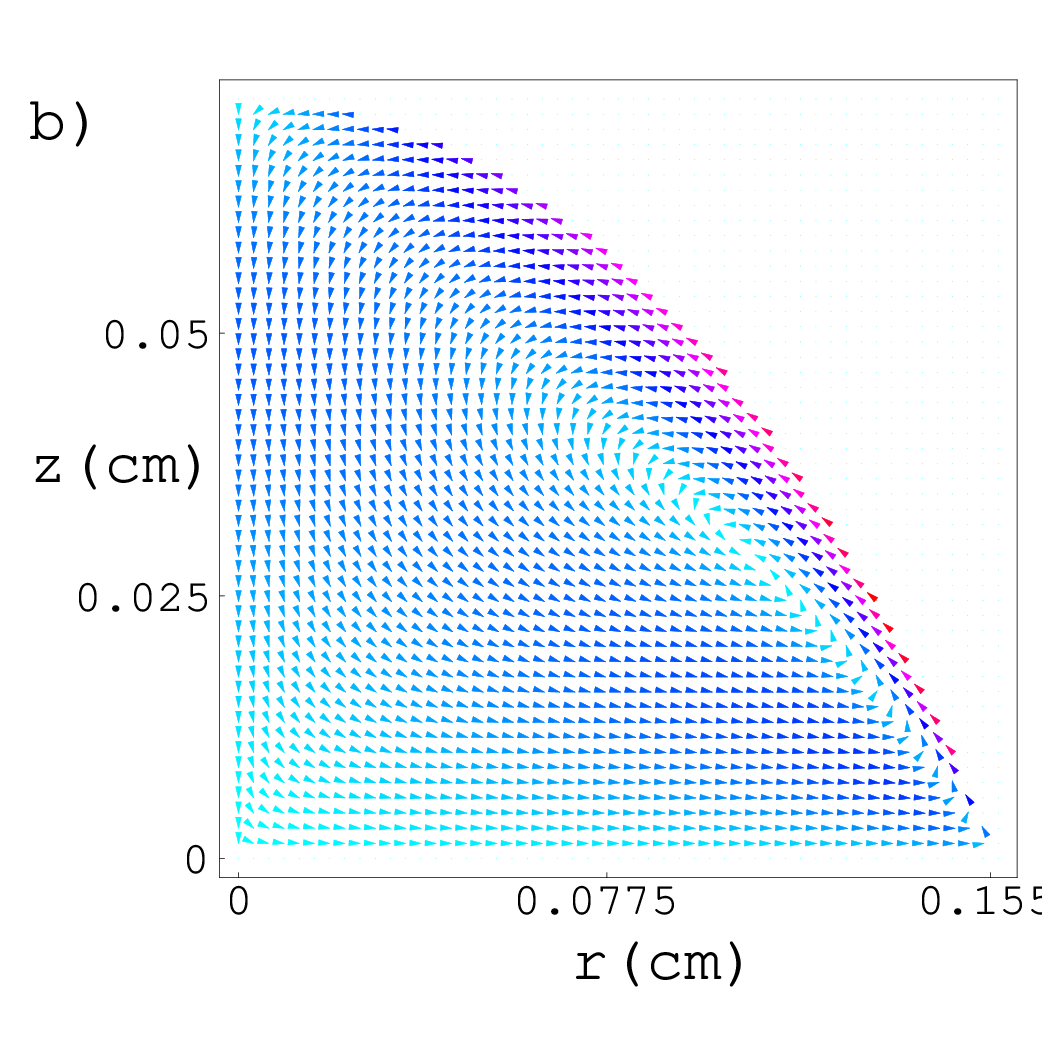}
\hspace{0.1cm}
\includegraphics[width=5.3cm]{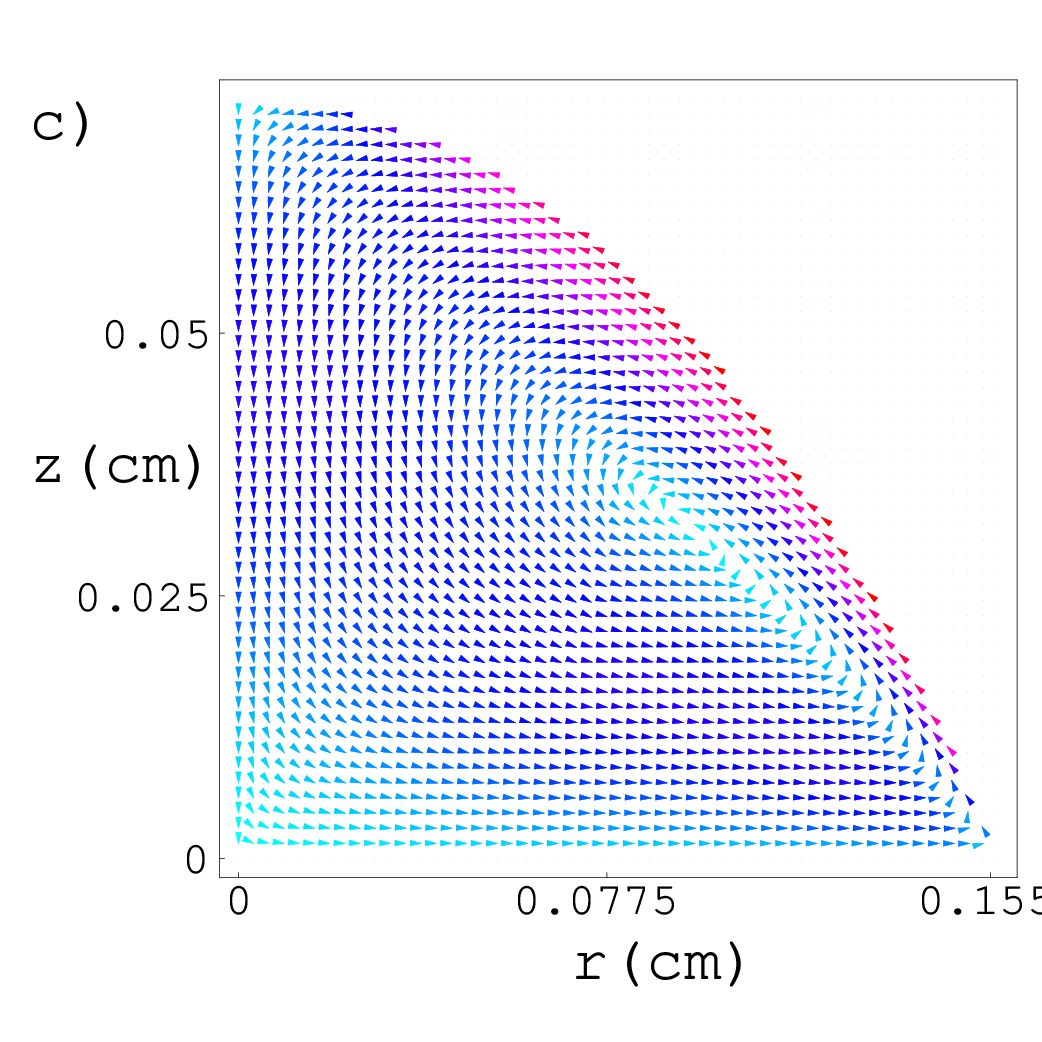}
\end{figure}
\begin{figure}[htb]
\caption{The surface velocity (cm/s) vs $r$ (cm) for the toluene droplet. 
Left panel: $\theta=50^\circ$, right panel: $\theta=20^\circ$.
Blue curve is the numerically obtained surface velocity, which shows
the ``bottleneck effect'' arising in the numerical results 
only when the Marangoni number exceeds 3000.
Green curve: the surface velocity in the heuristic extension of the lubrication approximation.
Purple curve: the surface velocity in the $n\tau$-description derived in Sec.~\ref{quadratic_ntau_Sec}.
}
\label{toluene_surface}
\vspace{0.5cm}
\includegraphics[width=8cm]{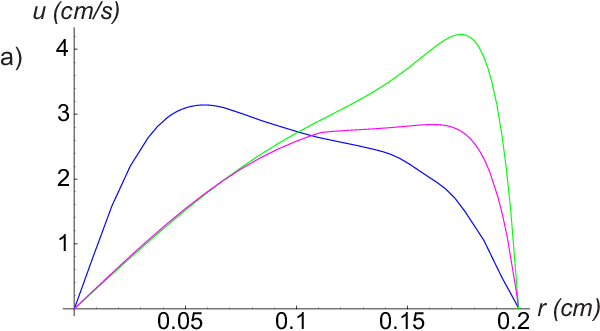}
\hspace{0.8cm}
\includegraphics[width=8cm]{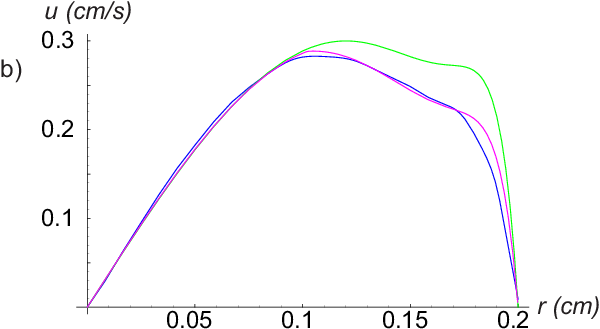}
\end{figure}
\begin{figure}[hbt]
\caption{Vector field plots of the velocity field containing single vortex
for the toluene droplet with $\theta=50^\circ$:
numerically obtained velocity field, which shows
the ``bottleneck effect'' arising in the numerical results 
only when the Marangoni number exceeds 3000,
the velocity field in the heuristic extension of the lubrication approximation
and in the $n\tau$-description.
}
\label{t4t_vect}
\vspace{0.2cm}
\includegraphics[trim=0.5cm 0.9cm 0 0.9cm,width=5.7cm]{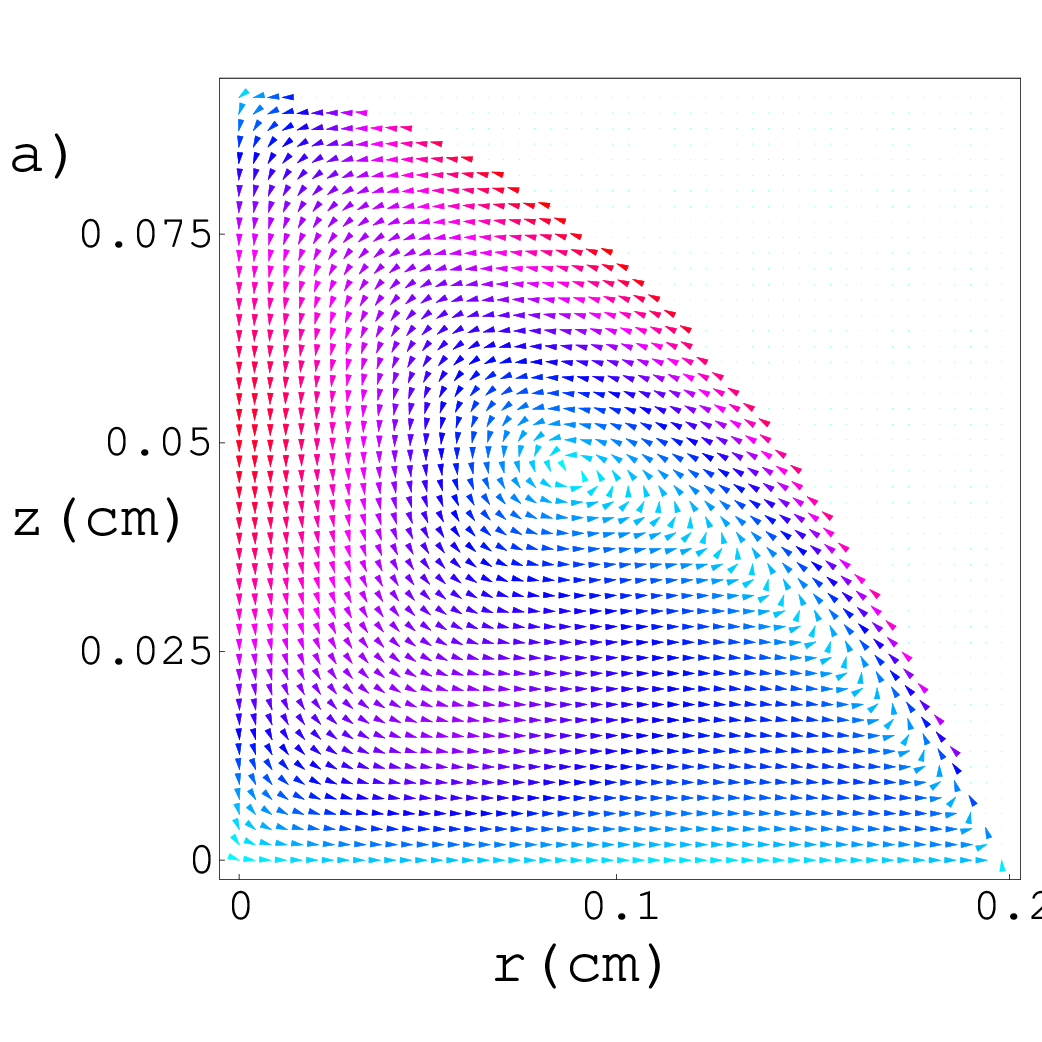}
\hspace{0.1cm}
\includegraphics[width=5.3cm]{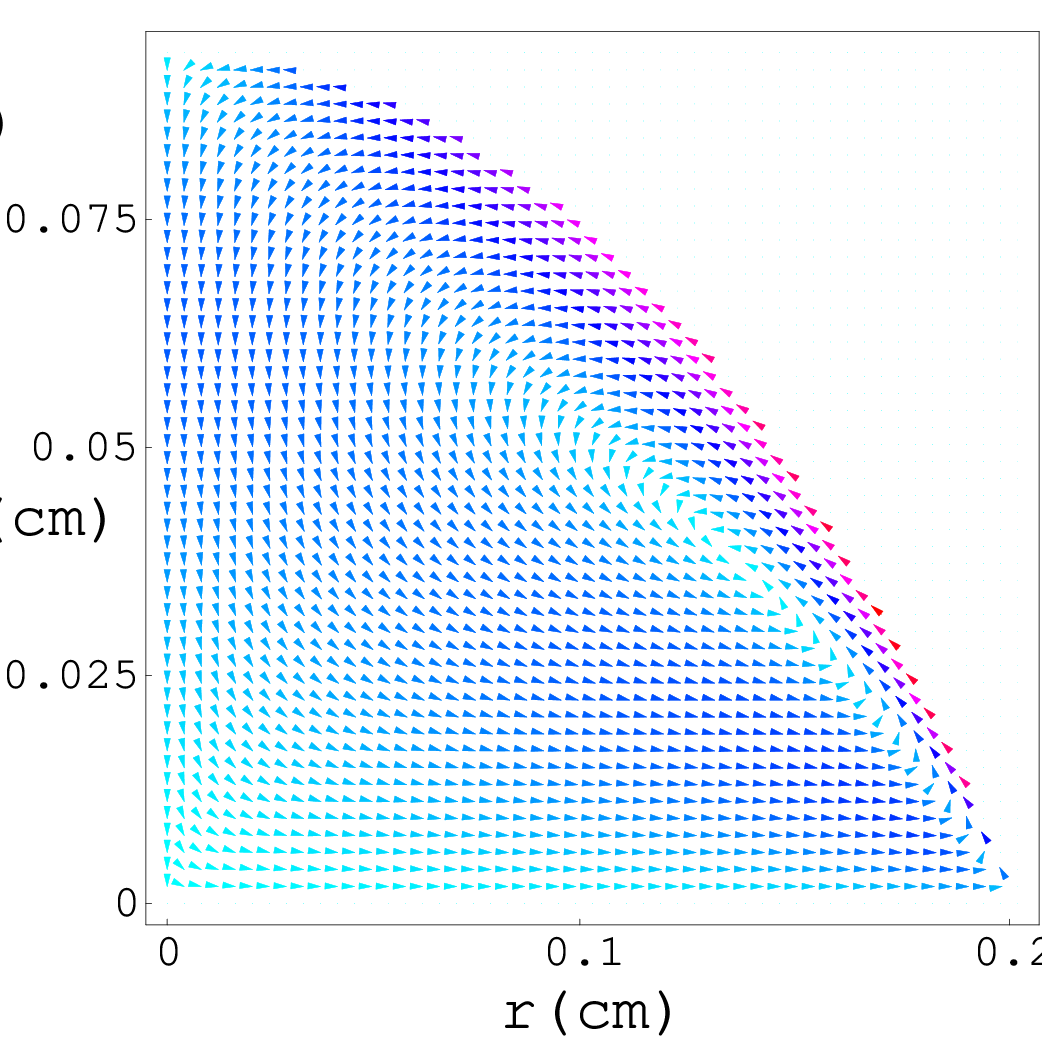}
\hspace{0.2cm}
\includegraphics[width=5.3cm]{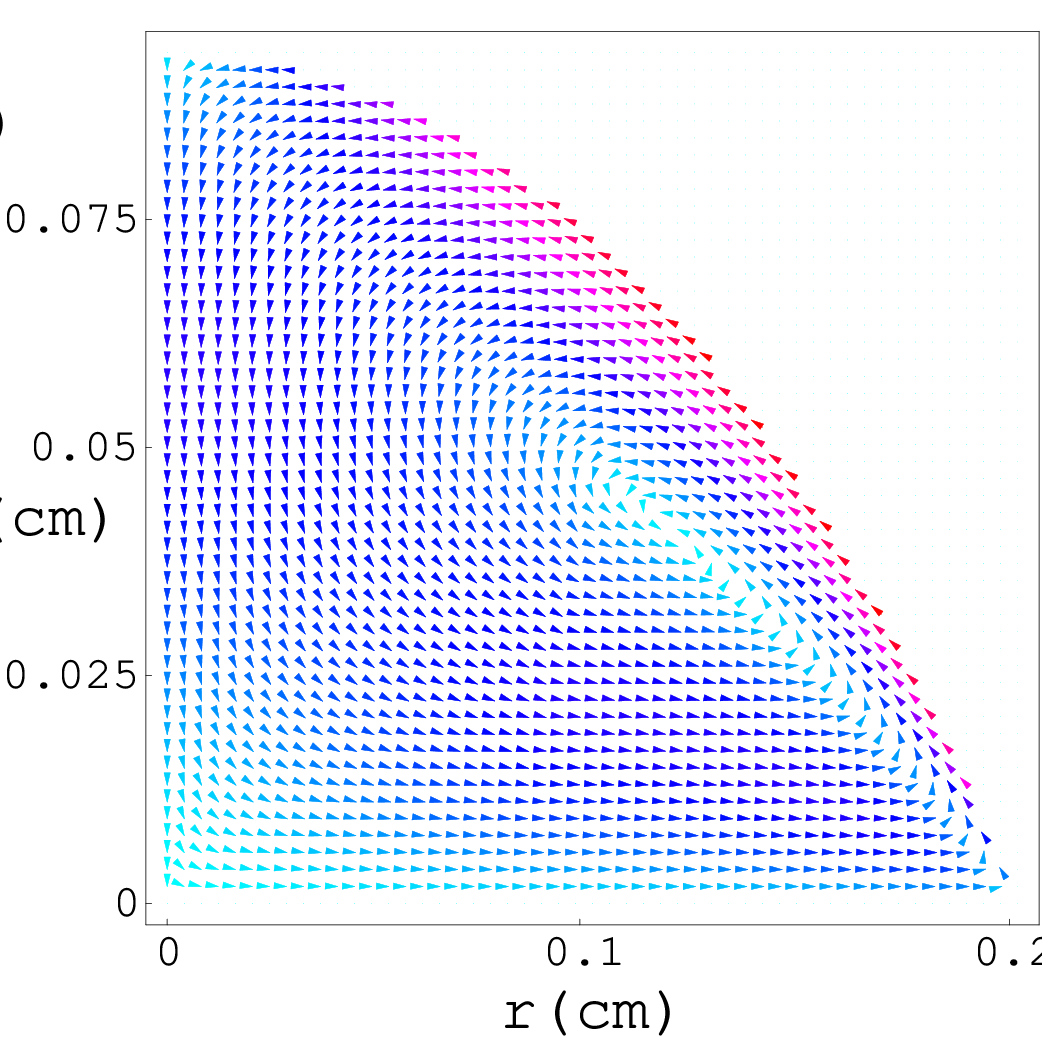}
\end{figure}

The numerical simulation for the fluid flows
was carried out with the method described in~\cite{Barash2009},
where the droplet surface was considered to be fixed,
the surface temperature was taken in accordance with
Table~\ref{TfitTable} and Eq.~(\ref{Tfit}),
and the heat convection was switched off. Although the inverse Stanton number for many
of the considered droplets is quite large, we switch off the heat convection,
since all the three approaches obviously do not take into account such effects.
Among the droplets considered, the inverse Stanton number is sufficiently small for a hexanol droplet 
with a contact angle $20^\circ$. It will also be small for droplets of a smaller size.
During the numerical calculation, the array of velocity values is obtained
from the stream function~\cite{Barash2009} using the relations
\begin{equation}
v_r(i,j)=\frac{\psi_{i,j+1}-\psi_{i,j}}{i h_x h_y},\qquad\qquad
v_z(i,j)=\frac{\psi_{i,j}-\psi_{i+1,j}}{(i+1/2) h_x h_x}.
\end{equation}
Therefore, we will estimate the deviation of the numerical velocity field
from the analytically obtained velocity field using the following mean-square deviations:
\begin{eqnarray}
\sigma_r&=&\frac{1}{N}\sqrt{\sum_{i,j=0}^{N-1}\left(v_r(i,j)-u_r\left(i h_x,\left(j+1/2\right) h_y\right)\right)^2},\\
\sigma_z&=&\frac{1}{N}\sqrt{\sum_{i,j=0}^{N-1}\left(v_z(i,j)-u_z\left(\left(i+1/2\right) h_x,j h_y\right)\right)^2}.
\end{eqnarray}
Here $N=200$ is the mesh size, $h_x=R/N$, $h_y=h/N$.
One more characteristic value of the velocity field 
is $u_{max}$, the absolute value of maximal velocity
at the surface of the droplet. 
Tables~\ref{TabVeloc1} and~\ref{TabVeloc2} 
show $u_{max}$ obtained with numerical simulation, 
with heuristic extension of the lubrication approximation,
and with the two versions of the $n\tau$-approximation
which employ $p(\phi)=2$ and $p(\phi)$ taken from Eq.~(\ref{TrialFunction}) correspondingly.
The tables also show the values of $\sigma_r$ and $\sigma_z$
for the analytical descriptions.
Table~\ref{TabVeloc2} contains the results for the droplet
of 2-propanol, where the size of the droplet is varied.
Also, table~\ref{TabVeloc2} contains the comparison of analytical 
and numerical results for the droplet of virtual liquid with variable viscosity,
where all other characteristics coincide with those of 2-propanol.

Table~\ref{TabVeloc1} and Figs.~\ref{hexanol_surface}b and~\ref{toluene_surface}b
show for the case $\theta<30^\circ$, when the droplets are relatively flat,
that all the three approximate analytical descriptions, 
including the lubrication approximation, agree well with
the numerically obtained velocity field, though the
$n\tau$-approach with Eq.~(\ref{TrialFunction}) is the most precise.

For large contact angles, Table~\ref{TabVeloc1} shows that
the accuracy of the 
$n\tau$-description exceeds that of the heuristic
extension of lubrication
approximation by a factor of about $1.5$, and also the $n\tau$-description
results in a much more precise value of the maximal surface velocity
$u_{max}$. The heuristic extension 
of the lubrication approximation still works within
$15$--$30$ per cent for droplets with large contact angles,
where the assumptions a), b) and c) of Sec~\ref{LubriSec}
are not justified.

The droplets under consideration have a wide spread of values of the
Marangoni number $\tilde{\mathrm{M_a}}=-\sigma'_Tc_V\Delta T \rho h/(\eta k)$,
starting from $10$ for the droplet of 1-hexanol, up to $3450$ for 
the toluene droplet. For droplets of toluene, propanol, octane and ethanol,
the droplet size is much smaller than the Marangoni cell
size~\cite{Pearson,Barash2009} on the flat fluid film containing 
the same liquid of the same height,
while for droplets of butanol and hexanol, the droplet size
and the Marangoni cell size are of the same order.

The comparison shows that the discrepancy between the numerical 
results and the analytical descriptions is considerably large only for
droplets with huge Marangoni numbers and, therefore, large contact angles.
Numerical results show that when the Marangoni number
exceeds $3000$, the ``bottleneck effect'' will take place. This means that 
the vortex center becomes sufficiently close to the symmetry axis due to
large Marangoni forces. This results
in substantial increase of downward velocities along the symmetry axis.
This is shown in Figs.~\ref{toluene_surface}a and~\ref{t4t_vect}a,
where the color scale for vector field plots is shown in Fig.~\ref{hexanol_surface}c
and the values $u_{max}$ are given in Table~\ref{TabVeloc1}.
Evidently, such an effect cannot be quantitatively
described without employing the Navier--Stokes equations
and without taking into account the convective heat transfer.
Also, it seems very probable
that for such huge Marangoni numbers, Marangoni forces
would rather destroy the axial symmetry of the droplet,
which would result in a more complicated three-dimensional 
velocity field.
We observed the ``bottleneck effect'' only for the large droplet
of toluene and for the large droplet of virtual liquid with
viscosity in $4$ or $8$ times smaller than that of 2-propanol
(see the fourth and fifth rows in Table~\ref{TabVeloc2}).
For other droplets there is no ``bottleneck effect''.
For example, Figs.~\ref{hexanol_surface} and~\ref{h4t_vect}
show the velocity field for the droplet of 1-hexanol.
The velocity fields for all other droplets with contact
angles $50^\circ$ and  $20^\circ$ are very similar
to those shown in Figs.~\ref{hexanol_surface} and~\ref{h4t_vect}.
Thus, the numerical results for axially symmetrical velocity field
demonstrate a large discrepancy with the analytical description
only when the Marangoni and the inverse Stanton number 
are sufficiently large, while the results are applicable
when ${\mathrm{St}}^{-1}\ll 1$.

\section{Conclusion}

A comparatively simple and detailed description of the single vortex fluid flows,
which ensures a good accuracy when the inverse Stanton number is much smaller
than unity, has been developed for an axially symmetrical evaporating
sessile drops of capillary size. 
The results have been tested and compared with the data of numerical simulations 
for droplets of various liquids and for a wide range of contact angles. 
The approach analytically addresses the boundary conditions and the mass balance 
equations making a simple assumption regarding tangential velocity field, 
without explicitly solving the Navier--Stokes equations. 
The results obtained signify that the boundary conditions and 
the mass balance equations dominate in the formation of the single vortex 
convection inside the droplet.

The author thanks R.G.~Larson and Wei Chen for useful discussions.
The simulations were partially carried out using facilities of
the Supercomputing Center of Lomonosov Moscow State University~\cite{Lomonosov}.
The results in Secs.~\ref{quadratic_ntau_Sec},~\ref{AnalysisSec} were supported
by the Russian Science Foundation project No.~14-21-00158.

\begin{table}[p]
\caption{The parameter values used in the calculations.}
\label{ParamTable}
\small
\hspace{-2.25cm}
\begin{tabular}{|c|l|l|l|l|l|l|l|l|l|}
\hline
&&&& toluene& ethanol & 2-propanol & octane & 1-butanol & 1-hexanol \\
\hline
\hline
Droplet 
& Initial temperature & $T_0$ & K & $293.15$ & $293.15$ & $293.15$ & $293.15$ & $293.15$ & $293.15$ \\
& Contact line radius & $R$ & cm   & $0.2$ & $0.154$ & $0.168$ & $0.175$ & $0.157$ & $0.155$ \\
\hline
\hline
Fluid
&Density & $\rho$ & g/cm$^3$    & $0.87$ & $0.789$ & $0.786$ & $0.703$ & $0.8098$ & $0.8136$ \\
&Molar mass & $\mu$ & g/mole & $92.14$ & $46.07$ & $60.1$ & $114.23$ & $74.122$ & $102.17$\\
&Thermal conductivity & $k$ & W/(cm$\cdot$ K) & $1.311\cdot 10^{-3}$ & $1.69\cdot 10^{-3}$ & $1.35\cdot 10^{-3}$ & $1.28\cdot 10^{-3}$ & $1.54\cdot 10^{-3}$ & $1.50\cdot 10^{-3}$  \\
&Heat capacity  & $c_p$ & J/(mole K)   & $156.7$ & $112.3$ & $156.5$ & $254.6$ & $177.2$ & $240.4$  \\
&Isochoric heat capacity & $c_v$ & J/(g K) & $1.286$ & $1.787$ & $1.931$ & $1.791$ & $1.83$ & $1.889$ \\ 
&Thermal diffusivity & $\kappa$ & cm$^2$/s & $8.86\cdot 10^{-4}$ & $8.79\cdot 10^{-4}$ & $6.60\cdot 10^{-4}$ & $8.17\cdot 10^{-4}$ & $7.95\cdot 10^{-4}$ & $7.84\cdot 10^{-4}$ \\
&Dynamic viscosity & $\eta$ &  g/(cm$\cdot$ s) & $5.6\cdot 10^{-3}$ & $1.074\cdot 10^{-2}$ & $2.038\cdot 10^{-2}$ & $5.08\cdot 10^{-3}$ & $2.544\cdot 10^{-2}$ & $4.578\cdot 10^{-2}$ \\
&Surface tension & $\sigma$ &  g/s$^2$  & $28.3049$ & $21.97$ & $20.93$ & $21.14$ & $24.93$ & $25.81$ \\
&$-\p($surface tension$)/\p T$ & $-\sigma'_T$ & g/(s$^2\cdot$ K) & $0.1189$ &$0.0832$ &$0.0788$ &$0.095$ &$0.0898$ &$0.08$ \\
&Latent heat of evap. & $L$ & J/g & $300.0$ & $918.6$  & $755.2$  & $363.2$  & $706.27$  & $603.0$  \\
\hline
\hline
Vapor
&Diffusion constant & $D$ & cm$^2$/s & $0.1449$ & $0.1181$ & $0.1013$ & $0.0616$ & $0.0861$ & $0.0621$ \\
&Saturated vapor density & $u_s$ & g/cm$^3$ & $1.27\cdot 10^{-4}$ & $1.46\cdot 10^{-4}$ & $1.38\cdot 10^{-4}$ & $8.57\cdot 10^{-5}$ & $2.76\cdot 10^{-5}$ & $6.55\cdot 10^{-6}$ \\
& Local evap. rate at apex & $J_0$ & g/(cm$^2\cdot$ s) & $9.2\cdot 10^{-5}$ & $1.12\cdot 10^{-4}$ & $8.2\cdot 10^{-5}$ & $2.9\cdot 10^{-5}$ & $1.55\cdot 10^{-5}$ & $2.7\cdot 10^{-6}$ \\
\hline
\end{tabular}
\end{table}

\begin{table}[p]
\caption{Fitting parameters for calculation of surface temperature.}
\label{TfitTable}
\begin{tabular}{|c|c|c|c|c|c|c|c|c|}
\hline
&$\theta$& $R$, cm &$\Delta T$, K&a&b&c\\
\hline
 1-butanol&   $50^{\circ}$ &0.157& 0.529506 & 0.3793   & 10 & 552.629 \\
 1-butanol&   $20^{\circ}$ &0.157& 0.154023 & 0.324516 & 15 & 1902.29 \\
 ethanol&     $50^{\circ}$ &0.154& 4.55357  & 0.379299 & 10 & 63.3781 \\
 ethanol&     $20^{\circ}$ &0.154& 1.32454  & 0.324514 & 15 & 220.321 \\
 octane&      $50^{\circ}$ &0.175& 0.727793 & 0.379299 & 10 & 401.793 \\
 octane&      $20^{\circ}$ &0.175& 0.2117   & 0.324512 & 15 & 1383.74 \\
 2-propanol & $50^{\circ}$ &0.168& 3.7995   & 0.379299 & 10 & 76.1549 \\
 2-propanol & $20^{\circ}$ &0.168& 1.1052   & 0.324514 & 15 & 264.246 \\
 toluene &    $50^{\circ}$ &0.2& 2.04597  & 0.379299 & 10 & 142.281   \\
 toluene &    $20^{\circ}$ &0.2& 0.595135 & 0.324515 & 15 & 491.577   \\
\hline
 1-hexanol&   $10^{\circ}$ &0.155& 0.010597 & 0.29475  & 24 & 27662.5 \\
 1-hexanol&   $15^{\circ}$ &0.155& 0.016353 & 0.337253 & 16 & 17925.4 \\
 1-hexanol&   $20^{\circ}$ &0.155& 0.023109 & 0.324521 & 15 & 12684.5 \\
 1-hexanol&   $25^{\circ}$ &0.155& 0.030568 & 0.317682 & 14 & 9589.09 \\
 1-hexanol&   $30^{\circ}$ &0.155& 0.038757 & 0.317114 & 13 & 7562.79 \\
 1-hexanol&   $35^{\circ}$ &0.155& 0.047704 & 0.323616 & 12 & 6144.19 \\
 1-hexanol&   $40^{\circ}$ &0.155& 0.057442 & 0.337952 & 11 & 5102.41 \\
 1-hexanol&   $45^{\circ}$ &0.155& 0.068007 & 0.3489   & 11 & 4309.59 \\
 1-hexanol&   $50^{\circ}$ &0.155& 0.079445 & 0.379306 & 10 & 3688.97 \\
 1-hexanol&   $55^{\circ}$ &0.155& 0.091807 & 0.403695 & 10 & 3192.11 \\
 1-hexanol&   $60^{\circ}$ &0.155& 0.105159 & 0.434731 & 10 & 2786.68 \\
\hline
 2-propanol & $50^{\circ}$ &0.1&  3.7995   &  0.379299 & 10 & 76.1549  \\
 2-propanol & $50^{\circ}$ &0.05& 3.7995   &  0.379299 & 10 & 76.1549  \\
 2-propanol & $50^{\circ}$ &0.02& 3.7995   &  0.379299 & 10 & 76.1549  \\
 2-propanol & $50^{\circ}$ &0.01& 3.7995   &  0.379299 & 10 & 76.1549  \\
\hline
\end{tabular}
\end{table}

\begin{table}[p]
\caption{Accuracy of the analytical descriptions. 
The values of $u_{max}$, $\sigma_r$ and $\sigma_z$ are in cm/s.}
\label{TabVeloc1}
\hspace*{-1.8cm}
\footnotesize
\begin{tabular}{|l|c||c|c|c|c||c|c|c|c|c|c|}
\hline
& $\theta$ & Numerical & Lubrication &  $n\tau$, $p=2$ & $n\tau$, (\ref{TrialFunction}) & \multicolumn{2}{|c|}{Lubrication} & \multicolumn{2}{|c|}{$n\tau$, $p=2$} & \multicolumn{2}{|c|}{ $n\tau$, (\ref{TrialFunction})} \\
\cline{3-12}
&  & $u_{max}$ & $u_{max}$ & $u_{max}$ & $u_{max}$ & $\sigma_r$ & $\sigma_z$ & $\sigma_r$ & $\sigma_z$ & $\sigma_r$ & $\sigma_z$ \\
\hline
1-butanol    & $50^{\circ}$  &  0.12     &    0.18    & 0.14   & 0.122 & 0.0098 & 0.0064 & 0.0092 & 0.0054 & 0.0077 & 0.0047  \\
1-butanol    & $20^{\circ}$  &  0.0123   &    0.0129  & 0.0127 & 0.0124& $8.7\cdot 10^{-4}$ & $1.9\cdot 10^{-4}$ & $8.5\cdot 10^{-4}$ & $1.9\cdot 10^{-4}$ & $8.3\cdot 10^{-4}$ & $1.8\cdot 10^{-4}$ \\
ethanol      & $50^{\circ}$  &  2.29     &    3.43    & 2.64   & 2.31  & 0.221  & 0.189  & 0.187  & 0.135  & 0.160 & 0.122 \\
ethanol      & $20^{\circ}$  &  0.232    &    0.244   & 0.240  & 0.234 & 0.017  & 0.0037 & 0.0162 & 0.0036 & 0.016 & 0.0034 \\
octane       & $50^{\circ}$  &  0.876    &  1.324     & 1.017  & 0.889 & 0.0812 & 0.065  & 0.070  & 0.046  & 0.059 & 0.040 \\
octane       & $20^{\circ}$  &  0.089    &  0.094     & 0.093  & 0.090 & 0.0064 & 0.0014 & 0.0063 & 0.0014 & 0.0061 & 0.0013 \\
2-propanol   & $50^{\circ}$  &  0.929    &  1.429     & 1.098  & 0.960 & 0.078  & 0.053  & 0.0716 & 0.0409 & 0.060 & 0.035 \\
2-propanol   & $20^{\circ}$  &  0.097    &  0.101     & 0.100  & 0.098 & 0.0069 & 0.0015 & 0.0068 & 0.0015 & 0.0066 & 0.0014 \\
toluene      & $50^{\circ}$  &  3.14     &  4.23      & 3.25   & 2.84  & 0.49   & 0.62   & 0.43   & 0.55   & 0.41 & 0.54  \\
toluene      & $20^{\circ}$  &  0.285    &  0.300      & 0.295  & 0.288 & 0.020  & 0.0046 & 0.020  & 0.0044 & 0.019 & 0.0042 \\
\hline
\hline
1-hexanol     & $10^{\circ}$   &  $2.2\cdot 10^{-4}$  & $2.2\cdot 10^{-4}$   & $2.2\cdot 10^{-4}$ & $2.2\cdot 10^{-4}$  & $1.5\cdot 10^{-5}$ & $1.5\cdot 10^{-6}$ & $1.5\cdot 10^{-5}$ & $1.5\cdot 10^{-6}$ & $1.4\cdot 10^{-5}$ & $1.5\cdot 10^{-6}$  \\
1-hexanol     & $15^{\circ}$   &  $4.7\cdot 10^{-4}$  & $4.9\cdot 10^{-4}$   & $4.8\cdot 10^{-4}$ & $4.7\cdot 10^{-4}$  & $3.4\cdot 10^{-5}$ & $5.5\cdot 10^{-6}$ & $3.4\cdot 10^{-5}$ & $5.4\cdot 10^{-6}$ & $3.3\cdot 10^{-5}$ & $5.2\cdot 10^{-6}$  \\
1-hexanol     & $20^{\circ}$   &  $9.1\cdot 10^{-4}$  & $9.6\cdot 10^{-4}$   & $9.4\cdot 10^{-4}$ & $9.2\cdot 10^{-4}$  & $6.5\cdot 10^{-5}$ & $1.4\cdot 10^{-5}$ & $6.4\cdot 10^{-5}$ & $1.4\cdot 10^{-5}$ & $6.2\cdot 10^{-5}$ & $1.3\cdot 10^{-5}$ \\
1-hexanol     & $25^{\circ}$   &  $1.54\cdot 10^{-3}$ & $1.64\cdot 10^{-3}$  & $1.6\cdot 10^{-3}$ & $1.57\cdot 10^{-3}$ & $1.1\cdot 10^{-4}$ & $3.1\cdot 10^{-5}$ & $1.1\cdot 10^{-4}$ & $3.0\cdot 10^{-5}$ & $1.1\cdot 10^{-4}$ & $2.9\cdot 10^{-5}$ \\
1-hexanol     & $30^{\circ}$   &  $2.4\cdot 10^{-3}$  & $2.6\cdot 10^{-3}$   & $2.5\cdot 10^{-3}$ & $2.4\cdot 10^{-3}$  & $1.8\cdot 10^{-4}$ & $6.0\cdot 10^{-5}$ & $1.7\cdot 10^{-4}$ & $5.6\cdot 10^{-5}$ & $1.6\cdot 10^{-4}$ & $5.3\cdot 10^{-5}$ \\
1-hexanol     & $35^{\circ}$   &  $3.5\cdot 10^{-3}$  & $4.0\cdot 10^{-3}$   & $3.7\cdot 10^{-3}$ & $3.6\cdot 10^{-3}$  & $2.6\cdot 10^{-4}$ & $1.1\cdot 10^{-4}$ & $2.5\cdot 10^{-4}$ & $9.8\cdot 10^{-5}$ & $2.3\cdot 10^{-4}$ & $9.2\cdot 10^{-5}$ \\
1-hexanol     & $40^{\circ}$   &  $4.8\cdot 10^{-3}$  & $6.2\cdot 10^{-3}$   & $5.3\cdot 10^{-3}$ & $4.9\cdot 10^{-3}$  & $3.8\cdot 10^{-4}$ & $1.8\cdot 10^{-4}$ & $3.6\cdot 10^{-4}$ & $1.6\cdot 10^{-4}$ & $3.3\cdot 10^{-4}$ & $1.5\cdot 10^{-4}$\\
1-hexanol     & $45^{\circ}$   &  $6.5\cdot 10^{-3}$  & $9.0\cdot 10^{-3}$   & $7.3\cdot 10^{-3}$ & $6.7\cdot 10^{-3}$  & $5.2\cdot 10^{-4}$ & $2.9\cdot 10^{-4}$ & $4.9\cdot 10^{-4}$ & $2.5\cdot 10^{-4}$ & $4.3\cdot 10^{-4}$ & $2.3\cdot 10^{-4}$ \\
1-hexanol     & $50^{\circ}$   &  $8.73\cdot 10^{-3}$ & $1.35\cdot 10^{-2}$  & $1.0\cdot 10^{-2}$ & $9.07\cdot 10^{-3}$ & $7.3\cdot 10^{-4}$ & $4.7\cdot 10^{-4}$ & $6.9\cdot 10^{-4}$ & $4.0\cdot 10^{-4}$ & $5.8\cdot 10^{-4}$ & $3.5\cdot 10^{-4}$ \\
1-hexanol     & $55^{\circ}$   &  $1.1\cdot 10^{-2}$  & $2.0\cdot 10^{-2}$   & $1.4\cdot 10^{-2}$ & $1.2\cdot 10^{-2}$  & $9.8\cdot 10^{-4}$ & $7.4\cdot 10^{-4}$ & $9.6\cdot 10^{-4}$ & $6.2\cdot 10^{-4}$ & $7.5\cdot 10^{-4}$ & $5.2\cdot 10^{-4}$ \\
1-hexanol     & $60^{\circ}$   &  $1.5\cdot 10^{-2}$  & $2.9\cdot 10^{-2}$   & $1.9\cdot 10^{-2}$ & $1.7\cdot 10^{-2}$  & $1.3\cdot 10^{-3}$ & $1.1\cdot 10^{-3}$ & $1.4\cdot 10^{-3}$ & $9.5\cdot 10^{-4}$ & $1.0\cdot 10^{-3}$ & $7.7\cdot 10^{-4}$ \\
\hline
\end{tabular}
\end{table}

\begin{table}[p]
\caption{Accuracy of the analytical descriptions. 
Dependence on the viscosity and on the contact line radius.
The values of $u_{max}$, $\sigma_r$ and $\sigma_z$ are in cm/s.}
\label{TabVeloc2}
\footnotesize
\hspace*{-1cm}
\begin{tabular}{|l|c|c|c||c|c|c|c||c|c|c|c|c|c|}
\hline
& $\theta$ & $R$, cm & $\eta/\eta_0$ & Numerical & Lubrication & $n\tau$, $p=2$ & $n\tau$, (\ref{TrialFunction}) & \multicolumn{2}{|c|}{Lubrication} & \multicolumn{2}{|c|}{$n\tau$, $p=2$} & \multicolumn{2}{|c|}{$n\tau$, (\ref{TrialFunction})}  \\
\cline{5-14}
& & &  & $u_{max}$ & $u_{max}$ & $u_{max}$ & $u_{max}$ & $\sigma_r$ & $\sigma_z$ & $\sigma_r$ & $\sigma_z$ & $\sigma_r$ & $\sigma_z$  \\
\hline
2-propanol    & $50^{\circ}$   &  0.168 &  2     &   0.46   &  0.71   & 0.55   &  0.48 & 0.039 & 0.025 & 0.036 & 0.021 & 0.030 & 0.018 \\
2-propanol    & $50^{\circ}$   &  0.168 &  1     &  0.929   &  1.429  & 1.098  & 0.960 & 0.078 & 0.053 & 0.072 & 0.041 & 0.060 & 0.035 \\
2-propanol    & $50^{\circ}$   &  0.168 &  0.5   &   1.90   &  2.86   & 2.20   &  1.92 & 0.182 & 0.153 & 0.154 & 0.109 & 0.131 & 0.097 \\
2-propanol    & $50^{\circ}$   &  0.168 &  0.25  &   4.31   &  5.72   & 4.39   &  3.84 & 0.70  & 0.90  & 0.622 & 0.807 & 0.59  & 0.79  \\
2-propanol    & $50^{\circ}$   &  0.168 &  0.125 &   8.23   &  11.43  & 8.78   & 7.68  & 1.70  & 2.35  & 1.567 & 2.190 & 1.49  & 2.16  \\
\hline
\hline
2-propanol    & $50^{\circ}$   &  0.1   &  1     &   0.93   &  1.43   & 1.098 &  0.96   & 0.078 & 0.052 & 0.072 & 0.041 & 0.060 & 0.035 \\
2-propanol    & $50^{\circ}$   &  0.05  &  1     &   0.92   &  1.43   & 1.098 &  0.96   & 0.077 & 0.051 & 0.072 & 0.042 & 0.060 & 0.036 \\
2-propanol    & $50^{\circ}$   &  0.02  &  1     &   0.92   &  1.43   & 1.098 &  0.96   & 0.077 & 0.050 & 0.072 & 0.042 & 0.061 & 0.037 \\
2-propanol    & $50^{\circ}$   &  0.01  &  1     &   0.92   &  1.42   & 1.098 &  0.96   & 0.077 & 0.050 & 0.072 & 0.043 & 0.061 & 0.037 \\
\hline
\end{tabular}
\end{table}


\begin{thebibliography}{99}

\frenchspacing

\bibitem{Maxwell1877}
J. C. Maxwell, 
``Diffusion''- Collected Scientific Papers,
Cambridge: Encyclopedia Britannica, 1877.

\bibitem{Langmuir1918}
I. Langmuir,
The evaporation of small spheres,
Phys.~Rev. \textbf{12}, No. 5, pp.~368--370 (1918).

\bibitem{Fuchs}
N.~A.~Fuchs, Evaporation and droplet growth in gaseous media
(Pergamon Press, Oxford, 1959).

\bibitem{Erbil} H.Y. Erbil, 
Evaporation of pure liquid sessile and spherical suspended drops: A review,
Adv. Colloid Int. Sci. {\bf 170}, 67 (2012).

\bibitem{Larson} R.G.~Larson, Transport and deposition patterns in drying
sessile droplets, AIChE Journal {\bf 60(5)}, 1538 (2014).

\bibitem{HuLarsonReverse}
H.~Hu, R.G.~Larson, Marangoni effect reverses coffee-ring depositions,
J. Phys. Chem. B {\bf 110}, 7090 (2006).

\bibitem{SavinoFico}
R.~Savino, S.~Fico, 
Transient Marangoni convection in hanging evaporating drops,
Phys.~Fluids {\bf 16}, 3738 (2004).

\bibitem{Thokchom}
A.K.~Thokchom, S.K.~Majumder, A.~Singh,
Internal fluid motion and particle transport in externally
heated sessile droplets,
AIChE Journal, {\bf 62(4)}, 1308 (2016).

\bibitem{Masoud} H. Masoud, J. D. Felske, 
Analytical solution for inviscid flow inside an evaporating sessile drop,
Phys. Rev. E {\bf 79}, 016301 (2009).

\bibitem{Tarasevich} Yu. Yu. Tarasevich,
Simple analytical model of capillary flow in an evaporating sessile drop,
Phys. Rev. E {\bf 71}, 027301 (2005).

\bibitem{Gelderblom} H. Gelderblom et.al.,
How water droplets evaporate on a superhydrophobic substrate,
Phys. Rev. E {\bf 83}, 026306 (2011).

\bibitem{Barash2009} L.Yu.~Barash, T.P.~Bigioni, V.M.~Vinokur, L.N.~Shchur,
Evaporation and fluid dynamics of a sessile drop of capillary size,
Phys. Rev. E {\bf 79}, 046301 (2009).

\bibitem{HuLarsonMarangoni} H.~Hu, R.~G.~Larson, 
Analysis of the Effects of Marangoni Stresses on the Microflow in an Evaporating Sessile Droplet,
Langmuir {\bf 21}, 3972 (2005).

\bibitem{HuLarsonMicrofluid} H.~Hu, R.~G.~Larson, 
Analysis of the Microfluid Flow in an Evaporating Sessile Droplet,
Langmuir {\bf 21}, 3963 (2005).

\bibitem{Deegan2000} R.~D.~Deegan et al., 
Contact line deposits in an evaporating drop,
Phys. Rev. E {\bf 62}, 756 (2000).

\bibitem{HuLarsonEvap} H.~Hu, R.~G.~Larson, 
Evaporation of a Sessile Droplet on a Substrate,
J. Phys. Chem. B {\bf 106}, 1334 (2002).

\bibitem{Lide} D.~R.~Lide, {\it CRC Handbook of Chemistry and Physics} (CRC Press, 2004).

\bibitem{deGennes} P.~G.~de Gennes, 
Wetting: statics and dynamics,
Rev. Mod. Phys. {\bf 57}, 827 (1985).

\bibitem{Bonn} D.~Bonn, J.~Eggers, J.~Indekeu, J.~Meunier, E.~Rolley, 
Wetting and spreading,
Rev. Mod. Phys. {\bf 81}, 739 (2009).

\bibitem{Huh} C.~Huh and L.~E.~Scriven, 
Hydrodynamic model of steady movement of a solid/liquid/fluid contact line,
J. Colloid Interface Sci. {\bf 35}, 85 (1971).

\bibitem{PetsiBurganos2008} A.~J.~Petsi, V.~N.~Burganos, 
Stokes flow inside an evaporating liquid line for any contact angle,
Phys. Rev. E {\bf 78}, 036324 (2008).

\bibitem{Rednikov} A.~Rednikov and P.~Colinet, 
Singularity-free description of moving contact lines for volatile liquids,
Phys. Rev. E {\bf 87}, 010401(R) (2013).

\bibitem{RednikovAPS} A.~Rednikov, 
Relaxation of contact-line singularities solely by the Kelvin effect 
and apparent contact angles for isothermal volatile liquids in contact with air,
\verb#http://meetings.aps.org/Meeting/DFD13/Session/G33.7#

\bibitem{Pearson} J.~R.~A.~Pearson, 
On convection cells induced by surface tension,
J. Fluid Mech. {\bf 4}, 489 (1958).

\bibitem{DuanBadam} F.~Duan, V.~K.~Badam, F.~Durst, C.~A.~Ward,
Thermocapillary transport of energy during water evaporation,
Phys. Rev. E {\bf 72}, 056303 (2005).

\bibitem{MurisicKondic} N.~Murisic and L.~Kondic,
On evaporation of sessile drops with moving contact lines,
J. Fluid Mech. {\bf 679}, 219 (2011).

\bibitem{Sobac}
B.~Sobac, D.~Brutin, 
Thermocapillary instabilities in an evaporating drop deposited 
onto a heated substrate, 
Phys. Fluids, {\bf 24}, 032103 (2012)

\bibitem{Carle}
F. Carle, B. Sobac, D. Brutin, 
Experimental evidence of the atmospheric convective transport contribution 
to sessile droplet evaporation, Appl. Phys. Lett. {\bf 102}, 061603 (2013).

\bibitem{Girard2011}
F. Girard, M. Antoni, K. Sefiane,
Use of IR thermography to investigate heated droplet evaporation
and contact line dynamics, Langmuir {\bf 27}, 6744 (2011)

\bibitem{Ristenpart}
W. D. Ristenpart, P. G. Kim, C. Domingues, J. Wan, H. A. Stone,
Phys. Rev. Lett. 99, 234502 (2007).

\bibitem{BarashSubstrate1}
L.Yu.~Barash, 
Dependence of the fluid convection in an evaporating sessile droplet on the thermal
conductivity of the substrate,
\verb#http://ccp2011.ornl.gov/pdf/Abstracts/Barash_Lev_9.1b_8.pdf#

\bibitem{BarashSubstrate2}
L.Yu.~Barash,
Dependence of fluid flows in an evaporating sessile droplet
on the characteristics of the substrate,
Int. J. Heat and Mass Transfer {\bf 84}, 419 (2015)

\bibitem{Zhang}
K.~Zhang, L.~Ma, X.~Xu, L.~Luo, D.~Guo,
Temperature distribution along the surface of evaporating droplets,
Phys. Rev. E {\bf 89}, 032404 (2014).

\bibitem{Semenov}
S.~Semenov, V.M.~Starov, R.G.~Rubio, M.G.~Velarde,
Computer simulations of evaporation of pinned sessile droplets: influence
of kinetic effects, Langmuir {\bf 28}, 15203 (2012)

\bibitem{Lomonosov}  Voevodin Vl.V., Zhumatiy S.A., Sobolev S.I., Antonov A.S., 
Bryzgalov P.A., Nikitenko D.A., Stefanov K.S., Voevodin Vad.V., 
Practice of "Lomonosov" Supercomputer, Open Systems J., Moscow: 
Open Systems Publ., 2012, no.7. (In Russian)

\end{thebibliography}
\end{document}